\documentclass[a4paper,11pt]{article}
\pdfoutput=1
\usepackage{jheppub}

\usepackage{tikz,xcolor,hyperref}
\usepackage{amsmath, amssymb, amsthm, graphicx, epsfig, fancyhdr,epsfig, slashed}
\usepackage[normalem]{ulem}
\usepackage{tikzsymbols}
\usepackage{natbib}
\usepackage{float}
\usepackage{subcaption}
\DeclareUnicodeCharacter{2212}{-}

\newcommand{\gs}{g_\star}
\newcommand{\gss}{g_{\star s}}
\newcommand{\Trh}{T_\text{rh}}

\newcommand{\Tmax}{T_\text{max}}

\newcommand{\DNeff}{\Delta N_\text{eff}}

\newcommand{\mdm}{m_\psi}

\begin{document}
\title{Implications of DESI for Dark Matter \& Cosmic Birefringence}
\author[a]{Basabendu Barman,}
\author[b,c]{and Sudhakantha Girmohanta}
\affiliation[a]{\,\,Department of Physics, School of Engineering and Sciences,
SRM University AP, Amaravati 522240, India}
\affiliation[b]{Tsung-Dao Lee Institute, Shanghai Jiao Tong University,
No.~1 Lisuo Road, Pudong New Area, Shanghai 201210, China}
\affiliation[c]{School of Physics and Astronomy, Shanghai Jiao Tong University,
800 Dongchuan Road, Shanghai 200240, China}
\emailAdd{basabendu.b@srmap.edu.in}
\emailAdd{sgirmohanta@sjtu.edu.cn}
\abstract{We explore an interacting dark matter (DM)–dark energy (DE) framework that naturally yields an effective dynamical DE equation of state crossing the phantom barrier at early times, as indicated by recent DESI data, while also accounting for the observed isotropic rotation of the cosmic microwave background (CMB) linear polarization. Within this unified framework, we also explain the DM relic abundance without introducing additional fields or couplings. Depending on the DE potential, we identify two viable scenarios: a superheavy freeze-in DM requiring a high reheating temperature, or a strongly interacting dark sector with a GeV–TeV scale thermal DM candidate.
}
\maketitle
\section{Introduction}
\label{sec:intro}
Measurements of Baryon Acoustic Oscillations (BAO) from the Dark Energy Spectroscopic Instrument (DESI)~\cite{DESI:2024mwx}, combined with type Ia Supernova observations from the Union3 compilation~\cite{SupernovaCosmologyProject:1998vns} and the Dark Energy Survey Year-5 data~\cite{DES:2024jxu}, suggest the possibility that dark energy (DE) may be evolving throughout cosmic history. Notably, the data favor a DE equation of state (EoS) $w < -1$ at high redshifts and $w > -1$ at present, indicating a so-called ``phantom crossing'' behavior~\cite{Hu:2004kh,Guo:2004fq}. The most recent DESI DR2 results~\cite{DESI:2025zgx,DESI:2025fii} appear to favor such a scenario, though the uncertainties are still substantial. Future measurements with improved precision will be crucial to confirm whether this crossing reflects a genuine feature of the dark energy dynamics. 

Constructing consistent models that yield an equation-of-state parameter $w<-1$ has proven to be theoretically challenging. Many existing approaches, such as phantom models~\cite{Caldwell:2003vq,Elizalde:2004mq}, invoke ghost-like degrees of freedom, which often lead to instabilities and other theoretical pathologies~\cite{Carroll:2003st,Cline:2003gs,Hsu:2004vr}. While scalar-tensor theories with non-minimal couplings to gravity, such as Brans--Dicke theory, can in principle reproduce $w<-1$ behavior~\cite{Boisseau:2000pr,Torres:2002pe,Capozziello:2002rd,Carroll:2004hc,Liddle:2003as}, stringent solar-system constraints render the Brans--Dicke scalar effectively inert, thereby forcing $w\approx-1$. A well-studied class of models is quintessence~\cite{Copeland:2006wr,PhysRevD.37.3406,Zlatev:1998tr,Bamba:2012cp}. In quintessence models, the evolution of the scalar field generally falls into two distinct categories: {\it freezing} models, where the field slows down over time and asymptotically approaches $w\to-1$, and {\it thawing} models, where the field remains initially frozen near $w\simeq -1$ and begins to evolve only recently. Although a variety of potentials, such as exponential, power-law, and others, have been extensively studied, they universally predict an equation-of-state parameter in the range $-1\leq w < 1/3$~\cite{Tsujikawa:2013fta,Bhattacharya:2024hep,Luu:2025fgw}. This range supports cosmic acceleration but does not extend into the phantom regime, making it challenging for such models to accommodate the behavior suggested by the DESI results~\cite{Wolf:2024eph,Ramadan:2024kmn,Bhattacharya:2024kxp}. Alternative proposals for achieving $w<-1$ include brane-world models~\cite{Sahni:2002dx}, quantum corrections~\cite{Onemli:2002hr}, pixelated Universe~\cite{Heckman:2024apk}, couplings between quintessence and moduli fields~\cite{Chung:2002xj}, non-minimally coupled quintessence field~\cite{Wolf:2024stt,Wolf:2025jed}, and photon–axion conversion mechanisms~\cite{Csaki:2004ha}. Moreover, the observed phantom-like behavior may also arise from a non-minimal coupling of DE to gravity~\cite{Carroll:2004hc}, or through interactions between the DE and dark matter (DM)~\cite{Huey:2004qv,Das:2005yj,Chakraborty:2025syu}\footnote{In~\cite{Berbig:2024aee} another possibility has been discussed, where a quintessence field possesses a non-zero initial velocity.}. Very recently, such models have also been tested using the gravitational-wave observations from LIGO and VIRGO~\cite{Yang:2019vni,Bachega:2019fki,Li:2019ajo}. Models of interacting DE and DM have also been shown to alleviate the $S_8$ tension, which is an apparent discrepancy between the amplitude of matter clustering inferred from the cosmic microwave background (CMB) and that observed at late times in weak lensing surveys~\cite{Poulin:2022sgp, Khoury:2025txd}. Models featuring DE–DM interactions could offer an appealing twofold resolution: they not only allow $w<−1$, and help to alleviate the $S_8$ tension, but may also hint at the origin of DM. Interacting DM-DE models have been studied in various contexts\footnote{In related models of DE conformally coupled with matter, dubbed as ``chameleon dark energy'', the interaction with local matter overdensities can effectively raise the local expansion rate, potentially resolving the Hubble tension without conflicting with large-scale observations~\cite{Cai:2021wgv} (see also Ref.~\cite{Karwal:2021vpk}).}, see, for example, Refs.~\cite{Damour:1990tw,Casas:1991ky,Bean:2001ys,Comelli:2003cv,Franca:2003zg,Chimento:2003iea,Amendola:1999er,Amendola:2000uh,Amendola:2003eq,Olivares:2005tb,Gubser:2004uf,Brax:2004qh,Copeland:2006wr,Bolotin:2013jpa,Andriot:2025los,Li:2024qso}, however, they lack a consistent picture of particle DM production.    

To shed light on the production mechanism of particle DM, one needs the knowledge of how the dark sector (DS) interacts within itself and with the visible sector. We take hints from the recently observed non-zero isotropic cosmic birefringence (ICB), characterized by the rotation of the polarization plane of CMB radiation~\cite{Minami:2020odp,Diego-Palazuelos:2022dsq,Eskilt:2022wav,Eskilt:2022cff, ACT:2025fju}, which could be explained by the in-flight interaction of CMB photons with the evolving DE background~\cite{Minami:2020odp,Fujita:2020ecn,Fujita:2020aqt,Takahashi:2020tqv,Obata:2021nql,Fung:2021wbz,Nakagawa:2021nme,Jain:2021shf,Choi:2021aze,Murai:2022zur,Tada:2024znt,Nakagawa:2025ejs}. Embedding this phenomenon within the interacting DE-DM framework then naturally induces a DM-Standard Model (SM) interaction, that could be linked to its observed relic density. Furthermore, as the interacting DM-DE framework comes with natural dynamics taking place within the DS itself~\cite{Khoury:2025txd}, it is natural to ask whether this is responsible for giving the DM its current abundance. We want to consider DM-genesis \textit{minimally}, where we do not introduce any additional fields or couplings other than what is already provided by this framework.

Motivated by these considerations, the present work aims to simultaneously address two key questions: (a) how to construct a DE model that yields an equation-of-state parameter $w<-1$ at low redshifts, and (b) how such a model can concurrently account for DM genesis within an interacting DE–DM framework, while also addressing the non-zero ICB angle. To this end, we build upon the model proposed in Ref.~\cite{Das:2005yj}, where a canonical scalar field, serving as the quintessence DE component, interacts with a fermionic DM. We extend this setup by introducing a dimension-5 operator coupling the scalar field to the electromagnetic (EM) field strength tensor. We consider two representative forms for the DE scalar potential: a polynomial potential and an axion-like potential. We give an ultraviolet (UV) completion for the model proposed in Ref.~\cite{Khoury:2025txd} and demonstrate how the EM coupling can be induced, which explains the observed ICB angle. Interestingly, the dynamics of the DS lead to very different DM mass scales in the two cases: while the polynomial potential supports $\gtrsim\mathcal{O}(10^{10})$ GeV DM (produced via freeze-in), the axionic scenario favors a GeV-TeV DM candidate (produced from DS freeze-out). We also outline the possibility of asymmetric DM production in the latter case, which requires additional extension of the model parameters, but could potentially give rise to DM self-interaction strength desirable from the explanation of diversity of rotation curves of spiral galaxies~\cite{Roberts:2024uyw}.  In both frameworks, the viable parameter space is found to be consistent with the observed DM relic abundance, recent DESI data, and the measured ICB angle.  

The paper is organized as follows. The underlying interaction, along with the dynamics of the interacting DM-DE model, is elaborated in Sec.~\ref{sec:framework}, in Sec.~\ref{sec:dm} we discuss DM production mechanisms, and finally conclude in Sec.~\ref{eq:concl}.
\section{The framework}
\label{sec:framework}
We introduce a scalar $\varphi$, which plays the role of evolving dark energy (DE) in the late Universe. Further, a Dirac fermion dark matter (DM), denoted $\psi$, is also assumed to interact with the DE as described by the following Lagrangian~\cite{Das:2005yj, Lin:2013sca, Girmohanta:2023ghm, Chakraborty:2025syu}
\begin{align}
    {\cal L} = \frac{1}{2} (\partial_\mu \varphi)^2 - V(\varphi) + i \bar \psi \gamma^\mu \partial_\mu \psi - m_{\rm} (\varphi) \bar\psi \psi + c_\gamma \left(\frac{\alpha_{\rm em}}{4 \pi}\right)\frac{\varphi}{f} F_{\mu \nu} \tilde{F}^{\mu \nu} \ ,
    \label{Eq:LDMDE}
\end{align}
where $V(\varphi)$ is the DE potential, $\alpha_{\rm em}$ is the EM fine-structure constant, $F^{\mu \nu}$ is the EM field strength tensor, $\tilde{F}^{\mu \nu} = (1/2) \epsilon^{\mu \nu \rho \sigma} F_{\rho \sigma}$ is its dual, and $f$ is the characteristic scale associated with $\varphi$. For instance, if $\varphi$ is a quintessence axion, $f$ plays the role of its decay constant. Notice we have introduced the anomalous EM coupling of $\varphi$ having Wilson coefficient $c_\gamma$ that could explain the ICB~\cite{Lin:2013sca}. The source of DM-DE interaction\footnote{
If there is no symmetry protecting the DE mass and potential shape, interactions with the DM can significantly raise the DE mass and destroy its potential shape, especially when there is a large hierarchy between the DE and DM masses~\cite{DAmico:2016jbm}.  However, if there is a remnant symmetry, such as a non-linearly realized shift symmetry of a broken scale/chiral invariance, that would protect the DE mass and potential shape. The scale invariance is suitable for a scalar (dilaton) like DE candidate~(see for \textit{e.g.,} Ref.~\cite{Brax:2004ym}), while the remnant shift symmetry can protect the pseudoscalar (axion) DE mass and potential against radiative instability.} is the $\varphi$ dependent DM mass, denoted as $m(\varphi)$ in Eq.~\eqref{Eq:LDMDE}. If $\varphi$ is a pseudo-scalar like an axion, associated with a dark QCD sector, then the anomalous coupling with $F^{\mu \nu}$ is natural, and can be induced if the hypercharge current is anomalous under the non-linearly realized shift symmetry of the axion. In this case, the scalar-like coupling with $\psi$ can appear as a result of finite density DM effects on the dark QCD condensate, provided $\psi$ is identified with the stable dark baryon DM~\cite{Khoury:2025txd}. If, on the other hand, $\varphi$ is a scalar, the required EM coupling can be generated if there are parity-violating interactions in the DS such that parity-odd and even states can mix. 
\subsection{Evolution of the DE and effective equation of state}
Let us briefly summarize the evolution of $\varphi$ as dictated by Eq.~\eqref{Eq:LDMDE}. We are mostly interested in the late time Universe, where the backreaction effect due to the EM coupling can be neglected on the evolution of $\varphi$. The Hubble parameter $H$ can be expressed as 
\begin{align}
    3 H^2 M_P^2 & = \rho_\varphi + \frac{\rho_{\rm DM}^{(0)}}{a^3} \frac{m(\varphi)}{m(\varphi_0)} + \frac{\rho_{\rm B}^{(0)}}{a^3} + \frac{\rho_{\rm rad}^{(0)}}{a^4} \ ,
    \label{Eq:Hgen}
\end{align}
 where $M_P$ is the reduced Planck mass, $a$ denotes the scale factor, the label `$0$' indicates the present-day value of the corresponding quantities, $\rho_{\rm DM}$, $\rho_{\rm B}$, and $\rho_{\rm rad}$ stand for the energy density stored in DM, baryons, and radiation component, while the DE energy density $\rho_\varphi$ is given as 
 \begin{align}
     \rho_\varphi = \frac{\dot{\varphi}^2}{2} + V(\varphi) \ ,
     \label{Eq:rhoPhi}
 \end{align}
 and the overdot denotes time derivative. Notice that due to non-trivial dependence of the DM mass on $\varphi$, the DM energy density scales as $\rho_{\rm DM}(a) \propto m(\varphi(a))/a^3$. This is the source of the effective equation of state for the DE being different from the non-interacting DE models~\cite{Das:2005yj}. We redefine dimensionless quantities $\eta(a)$, $\theta(a)$, $\tilde{f}$ for the quantities of interest as follows
 \begin{align}
     H (a) & \equiv \eta(a) H_0 \quad ; \quad \varphi(a) \equiv \theta(a) f  \quad  ; \quad f \equiv \tilde{f} M_P \quad ; \quad  V(\varphi) \equiv \widetilde{V} (\varphi) \rho_c^{(0)} \ ,
     \label{Eq:Dimless}
 \end{align}
where $H_0$ represents the Hubble rate at present. Dividing Eq.~\eqref{Eq:Hgen} by the critical energy density today, \textit{i.e.}, by $\rho_c^{(0)} = 3 M_P^2 H_0^2$ , one can evaluate the Hubble rate parametrically as a function of $\theta(a)$ as follows 
 \begin{align}
\eta(a) = \left [ \left( \widetilde{V} (\varphi) + \frac{\Omega_{\rm DM}^{(0)}}{a^3} \frac{m(\varphi)}{m(\varphi_0)} + \frac{\Omega_{\rm B}^{(0)}}{a^3} + \frac{\Omega_{\rm rad}^{(0)}}{a^4} \right) \left\{ 1-\frac{a^2}{6} \tilde{f}^2 \left(\frac{d \theta}{da}\right)^2 \right\}^{-1} \right]^{1/2} \ ,
\label{Eq:Hsp}
\end{align}
where $\Omega_i^{(0)}\equiv \rho_i^{(0)}/\rho_c^{(0)}$, for a given species $i = {\rm B,\,DM,\,rad}$. Neglecting the backreaction of radiation density, the equation of motion (EoM) for the DE reads
\begin{align}
    \ddot{\varphi} + 3 H \dot{\varphi} + \frac{d V_{\rm eff}}{d\varphi} = 0 \ ,
    \label{Eq:phiEOM1}
\end{align}
where the effective potential for the DE receives contributions from the DM density and is given by
\begin{equation}
    V_{\rm eff} (\varphi) = V(\varphi) + \frac{\rho_{\rm DM}^{(0)}}{a^3} \frac{m(\varphi)}{m(\varphi_0)}\,.
    \label{Eq:Veff}
\end{equation}
Note that here we do not consider any explicit DE-baryon interaction. In the early Universe, as the DM density is much larger than the present-day DE density, Eq.~\eqref{Eq:Veff} implies that the contribution of DM to the effective potential is more important.  However, as the Universe expands, the DM is diluted, and the DE bare potential starts to play an important role in the evolution of $\varphi$. Simplifying the EoM, one arrives at
\begin{align}
    \tilde{f}^2 \eta a \left[ \frac{d}{da}\left( \eta a \frac{d\theta}{da} \right) + 3 \eta  \frac{d\theta}{da} \right] + 3 \frac{d \widetilde{V} (\theta) }{d\theta} + 3 \frac{ \Omega_{\rm DM}^{(0)}}{a^3} \frac{m'(\theta)}{m(\theta_0)} = 0 \ ,
    \label{Eq:phiEoM2}
\end{align}
where $m'(\theta) \equiv dm(\theta)/d\theta$. Eqs.~\eqref{Eq:Hsp} and~\eqref{Eq:phiEoM2} should be solved consistently as a boundary value problem for a given $\theta_0$. 

If an observer treats the DM as a constant mass fluid, the effective DE energy density due to the non-trivial DM energy density evolution can be read off from Eq.~\eqref{Eq:Hgen}, and is given as 
\begin{align}
\rho_{\rm DE}^{\rm eff} = \frac{\rho_{\rm DM}^{(0)}}{a^3} \left[ \frac{m(\varphi)}{m(\varphi_0)} - 1 \right] + \rho_\varphi \ .
\label{Eq:DEeff}
\end{align}
One needs to satisfy the following condition to explain the DE abundance today $(a=1)$, 
\begin{equation}
   \left[ \frac{\tilde{f}^2}{6}\left( \frac{d \theta}{da}\right)^2 + \widetilde{V} (\theta) \right]_{a=1} = \Omega_{\rm DE}^{(0)} \ .
\end{equation}
Now, taking the time derivative of Eq.~\eqref{Eq:DEeff}, and using Eq.~\eqref{Eq:phiEOM1}, one gets the following expression for the effective equation of state parameter for the DE, denoted as $w_{\rm eff}$, namely,
\begin{align}
 \frac{d{\rho}^{\rm eff}_{\rm DE}}{dt} + 3 H (1+ w_{\rm eff}) \rho_{\rm DE} = 0  \ ,
\end{align}
where 
\begin{equation}
     w_{\rm eff}= \frac{w_\varphi}{1+\frac{\rho_{\rm DM}^0}{a^3 \rho_\varphi} \left[ \frac{m(\theta)}{m(\theta_0)}-1 \right]} \ ,
     \label{Eq:weff}
\end{equation}
and the present-day EoS $w_{\varphi}$ is given by  
\begin{equation}
    w_\varphi =  \frac{\dot{\varphi^2}/2-V(\varphi)}{\dot{\varphi^2}/2+V(\varphi)} \ .
\end{equation}
Note from Eq.~\eqref{Eq:weff} that $w_{\rm eff}$ can cross the phantom barrier (at a certain scale factor, which we will show in a moment),  \textit{i.e.}, $w_{\rm eff} < -1$, if $m(\theta_0)>m(\theta)$, even though $w_{\varphi} \geq -1$. Therefore, treating the EoS as an effective parameter, one can see that there is no unphysical behavior even if $w_{\rm eff} < -1$. This is a desirable feature in light of the current DESI results. One can recast Eq.~\eqref{Eq:weff} in terms of the redefined variables as, 
\begin{equation}
    w_{\rm eff}(a) = \frac{\frac{\tilde{f}^2}{6} (a\,\eta\,\theta'(a))^2- \widetilde{V}(\theta)}{\frac{\tilde{f}^2}{6} (a\,\eta\,\theta'(a))^2 + \widetilde{V}(\theta) + \frac{\Omega_{\rm DM}^{(0)}}{a^3} \left( \frac{m(\theta)}{m(\theta_0)}-1 \right)}\,,
    \label{Eq:weffExplicit}
\end{equation}
where $\theta'(a)\equiv d \theta/da$. One can consistently solve Eqs.~\eqref{Eq:Hsp},~\eqref{Eq:phiEoM2} for some given $V(\theta), \  m(\theta)$ for a specific choice of $\tilde{f}, \theta_0$ by adjusting the initial conditions, and obtain the effective EoS for DE as a function of the scale factor or redshift utilizing~\eqref{Eq:weffExplicit}, and fit it with the DESI results. We will consider some concrete examples, and elaborate on the behavior of $w_{\rm eff}$. Note from Eq.~\eqref{Eq:weffExplicit} that, today $w_{\rm eff}$ approaches $-1$ if $\tilde{f} \ll 1$. 
\subsubsection*{Isotropic cosmic birefringence}
The phenomenon of cosmic birefringence~\cite{Carroll:1989vb,Carroll:1991zs,Harari:1992ea} refers to the rotation of the polarization plane of the CMB photons as they travel through space. The net rotation angle from the last scattering surface to the present is termed the ICB angle, denoted by $\beta$. Analysis of CMB polarization data has indicated a non-zero rotation angle~\cite{Minami:2020odp,Diego-Palazuelos:2022dsq,Eskilt:2022wav,Eskilt:2022cff}: $\beta = 0.34^\circ \pm 0.09^\circ$. Additionally, recent data from the Atacama Cosmology Telescope (ACT) further supports this result~\cite{ACT:2025fju}: $\beta = 0.20^\circ \pm 0.08^\circ$, 
strengthening the hypothesis that the ICB could be a genuine physical effect. Within the SM and its extensions, particularly those involving higher-dimensional operators built solely from SM fields, there is no satisfactory explanation for this phenomenon~\cite{Nakai:2023zdr}. Therefore, the observed ICB may suggest a parity-violating interaction involving a new light field beyond the SM. A nonzero isotropic birefringence angle $\beta$ can arise when the massive scalar field $\varphi$ is coupled to the SM photon field strength tensor $F_{\mu\nu}$ through an interaction term of the form $\propto(\varphi/f)\,F\,\tilde{F}$ (as in Eq.~\eqref{Eq:LDMDE}). In this framework, a nonvanishing difference $\Delta \varphi = \varphi_0 - \varphi_{\text{LSS}}$,
with $\varphi_0$, $\varphi_{\text{LSS}}$ being the values of the field today and at the last scattering surface (LSS), respectively, can account for the observed non-zero birefringence angle~\cite{Minami:2020odp,Fujita:2020ecn,Fujita:2020aqt,Takahashi:2020tqv,Obata:2021nql,Fung:2021wbz,Nakagawa:2021nme,Jain:2021shf,Choi:2021aze,Murai:2022zur,Tada:2024znt,Nakagawa:2025ejs} (also see Ref.~\cite{Komatsu:2022nvu} for a review). In particular, for the present framework~\cite{Takahashi:2020tqv,Choi:2021aze},
\begin{equation}
|c_\gamma| \simeq 4\,\left(\frac{1}{\Delta\theta}\right)\, \left(\frac{\beta}{0.261} \right)\,,
\end{equation}
considering the central value of $\beta=0.261$~\cite{Lin:2025gne}. Here, $\Delta\varphi$ is determined by the underlying potential governing DE, which, as previously discussed, also sets the effective DE equation of state we aim to constrain using DESI data. In contrast, $c_\gamma$, as we will elaborate in Sec.~\ref{sec:dm}, controls the present-day abundance of the DM $\psi$. Thus, our framework offers a unified explanation for the DESI observations, the ICB angle, and the origin of DM. 
\subsection{Polynomial potential}
\label{sec:poly}
A large class of models can be represented by the following effective parameters~\cite{Das:2005yj, Smith:2025grk, Chakraborty:2025syu}
\begin{equation}
    V(\theta) = \Lambda^4 \left( \frac{1}{\theta}\right)^\alpha \quad ; \quad m(\theta) = m_{\psi} \exp \left[ \kappa (\theta-\theta_0) \right] \ ,
    \label{Eq:poly}
\end{equation}
where $\Lambda \simeq {\cal O}(\rm meV)$ sets the DE scale today, $m_{\psi}$ is the present DM mass, and the dimensionless parameters $\alpha$, $\kappa$ control the DE potential and DM-DE interaction strength, respectively. Although $V(\varphi)$ is non-analytic at $\varphi=0$, it should be regarded only as an effective potential at the low energy theory. In the limit $\kappa \to 0$, one recovers a non-interacting DE model, while for $\alpha \to 0$, the DE becomes equivalent to a cosmological constant.  In order to account for dynamical DE, the DE field $\varphi$ needs to be very light, which gives rise to an additional long-range force between the DM particles. In
the linearized limit, this behaves as a scalar correction to gravity, and its strength relative
to the Newtonian potential is given by $V_{\rm DM-DM}/V_{\rm N}\simeq \kappa^2$~\cite{DAmico:2016jbm}. It is then possible to put an upper bound on $\kappa$ from the violation of the weak equivalence principle, requiring $|\kappa|\lesssim 1/3$. Following~\cite{Das:2005yj}, a combination of mass-to-light ratios in the Local Group, rotation curves of galaxies in clusters, and dynamics of rich clusters yields the upper bound $\kappa\lesssim 0.8$, for $f=M_P$.
\begin{figure}[htb!]
    \centering    
    \includegraphics[scale=0.3]{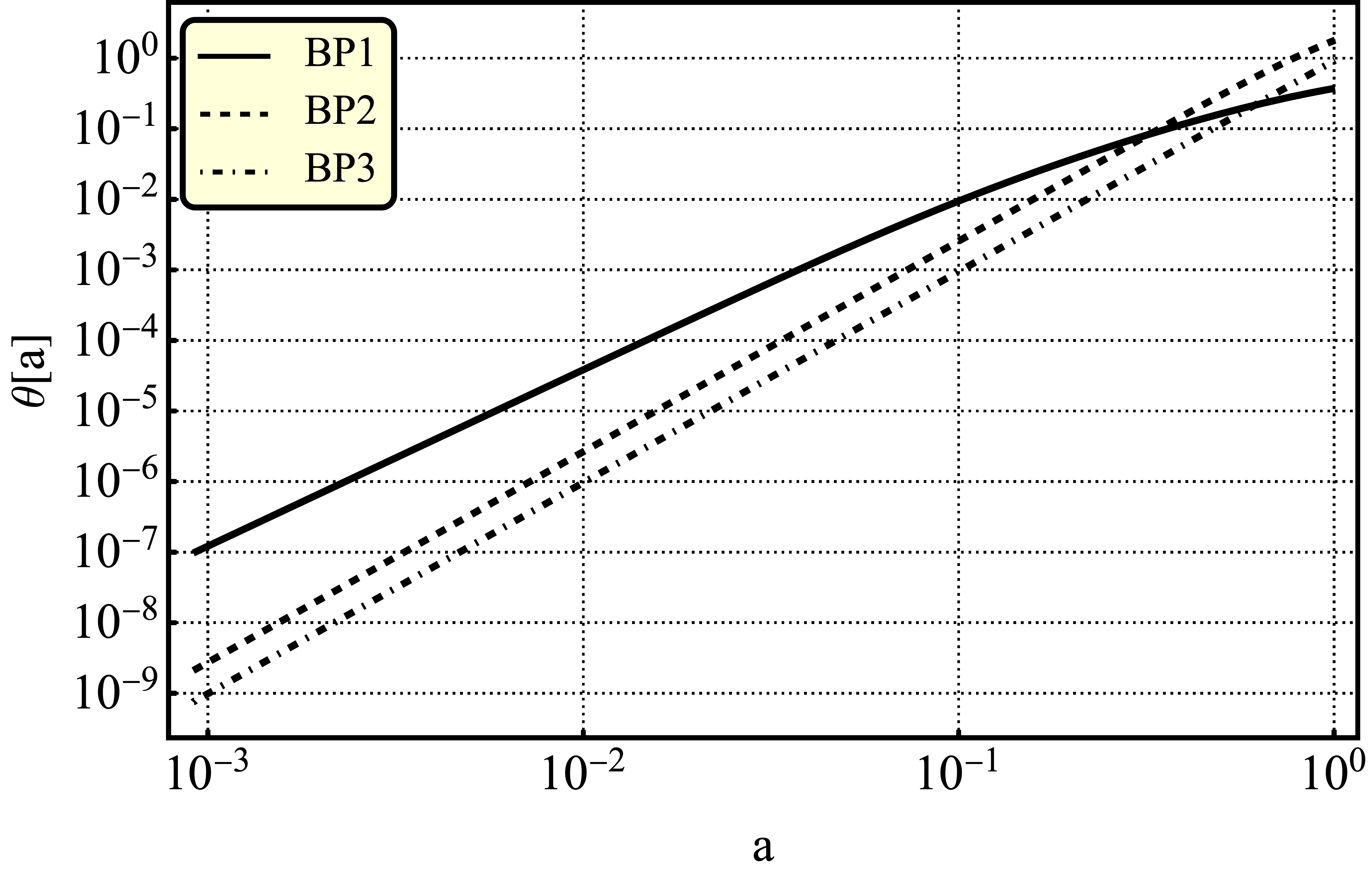}~
    \includegraphics[scale=0.32]{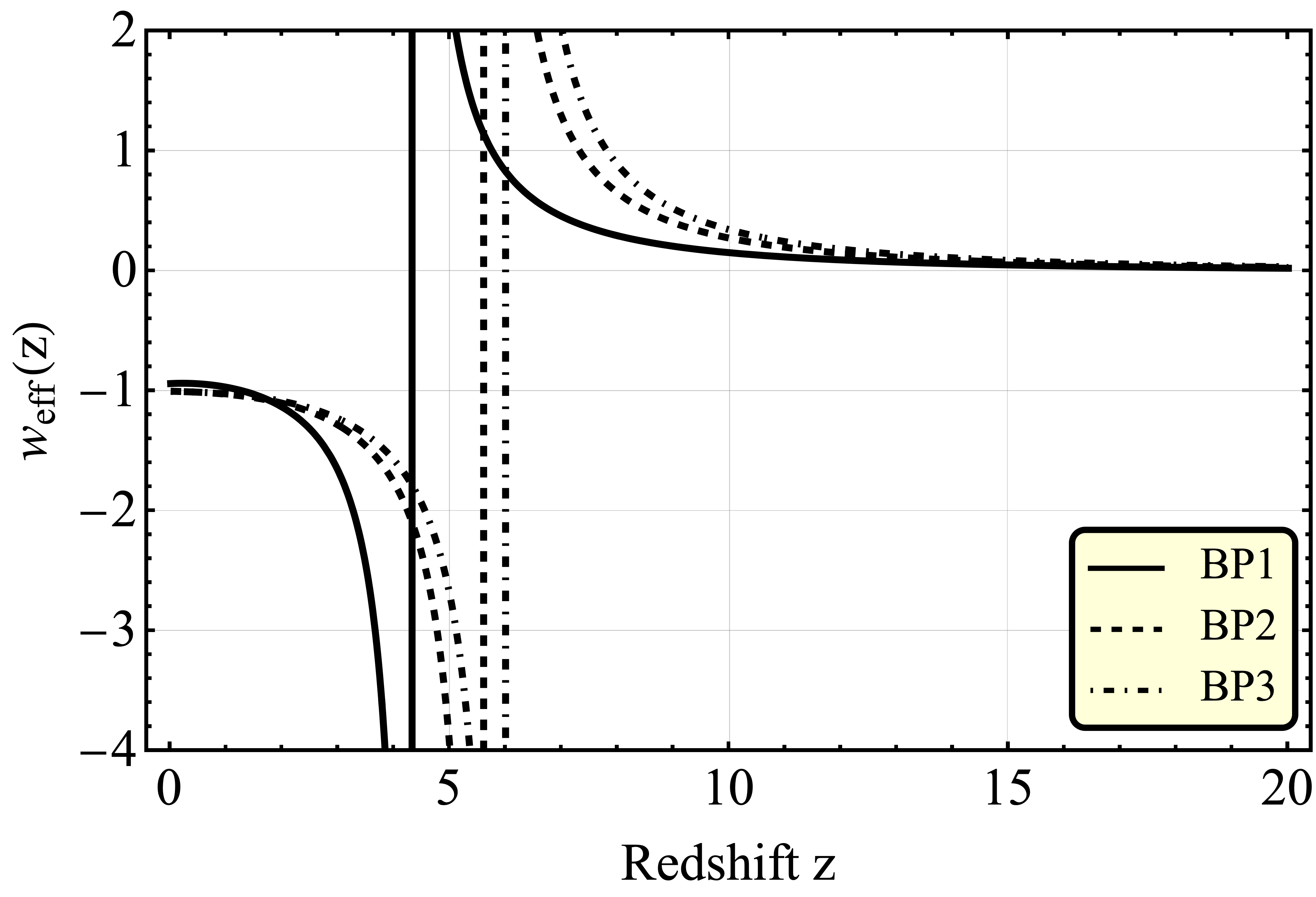}\\[10pt]
    \includegraphics[scale=0.4]{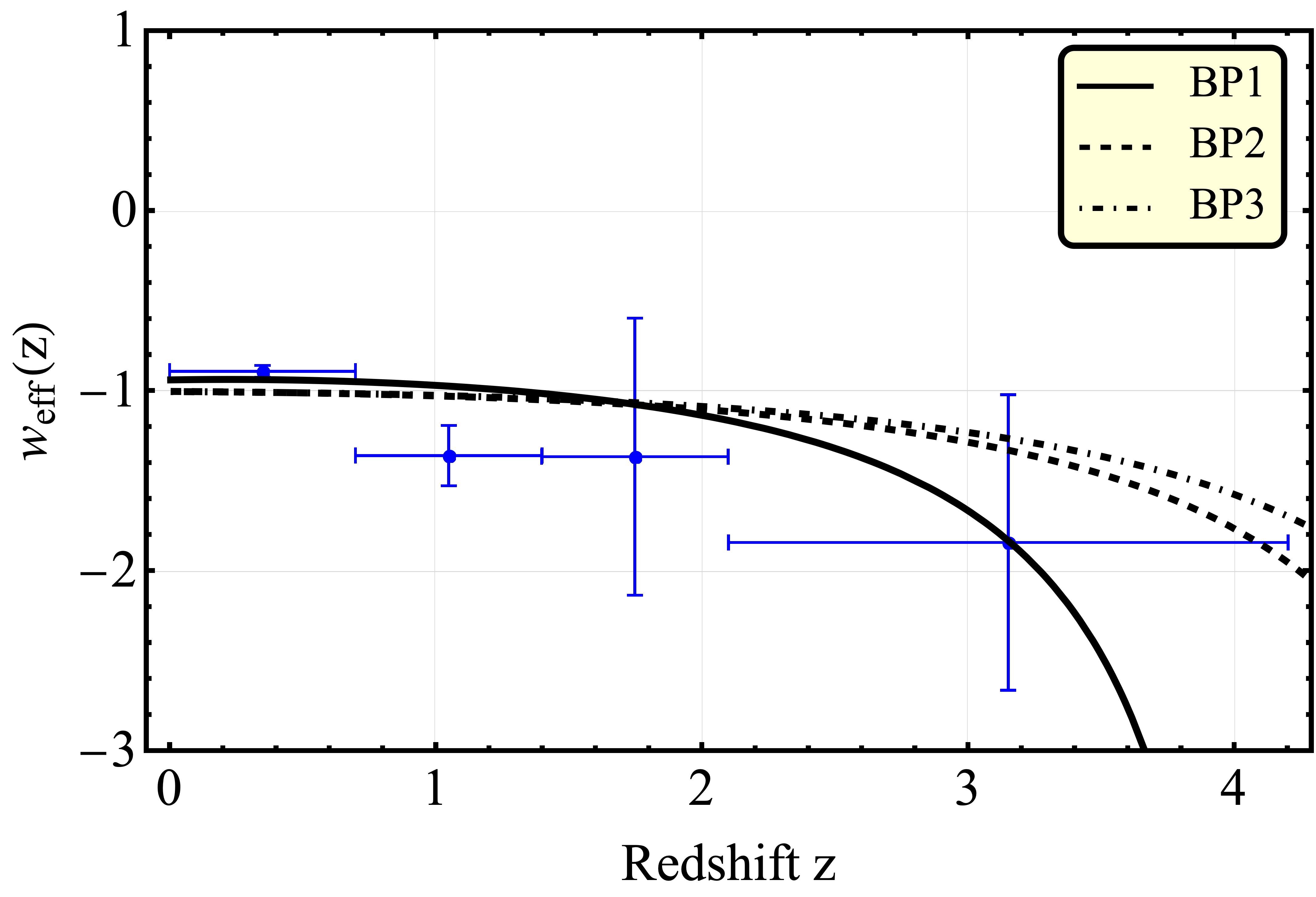}
    \caption{{\it Top left}: Evolution of the field as a function of scale factor $a$ for the polynomial potential. {\it Top right:} Nature of the $w_{\rm eff}$ as a function of redshift extrapolated to high redshift. It diverges momentarily, becomes positive, and approaches zero as one extrapolates to higher redshift.  {\it Bottom:} 
    Fit of the effective equation of state ($w_{\rm eff}$) as a function of redshift to the DESI data (blue dots) with binned error bars. All relevant parameters are tabulated in Tab.~\ref{tab:BP}.}
    \label{fig:fitPoly}
\end{figure}

In Fig.~\ref{fig:fitPoly} we show the evolution of the field and corresponding EoS for the DE for different choices of parameters. Note that the scale factor $a(t)$ and redshift $z$ are related via $1+z=a_0/a(t)\equiv 1/a(t)$, as we consider the present-day scale factor to be unity. Thus, a larger redshift corresponds to a smaller scale-factor, {\it i.e.,} an earlier time in the Universe. While solving the equation of motion for $\theta$, we have placed it at the minimum of the effective potential in Eq.~\eqref{Eq:Veff} initially. Then, the field smoothly tracks the evolving effective minimum of the potential with the expansion of the Universe. We have checked that if one places the field away from the effective minima at the starting position, it starts to oscillate around the effective minimum, soon reaches the minimum, and tracks the evolving minimum adiabatically thereafter. The effective EoS of DE at the late-time Universe and  the field excursion relevant for the ICB is effectively insensitive to the initial field position in the polynomial-type potential. We show the evolution of $\theta$ from the epoch of the last scattering (LSS), which takes place at $z\approx 1100\,(a\approx 10^{-3})$, till today, for the benchmark values of the parameters tabulated in Tab.~\ref{tab:BP}.
\begin{table}[htb!]
\centering
\begin{tabular}{|c|c|c|c|c|c|c|} 
\hline 
\rule{0pt}{2.5ex} Benchmarks & $\alpha$ & $\kappa$ & $\tilde{f}$ & $\Lambda$ [meV] & $\Delta \theta$ & $|c_\gamma|$ 
\\
\hline\hline
BP1 & 0.2 & 0.08  & 1 & 2.59 & 0.39 & $8-12$
\\
\hline
BP2 & $5\times 10^{-3}$ & $5\times 10^{-3}$ & $10^{-2}$ & 2.74 & 1.86 & $2 - 3$\\
\hline
BP3 & $3\times 10^{-3}$ & $8\times 10^{-3}$ & $10^{-3}$ & $2.74$ & $0.97$ & $3-5$ \\
\hline\hline
\end{tabular}
\caption{Benchmark points chosen for Fig.~\ref{fig:fitPoly}, for the polynomial potential scenario. In the last two columns, we have included the resulting field excursion $\Delta \theta$ and corresponding $c_\gamma$, compatible with a 1$\sigma$ value of the ICB angle $\beta= (0.261 \pm 0.061)^\circ$~\cite{Lin:2025gne}.} 
\label{tab:BP}
\end{table}
As one can see from the top right panel, the effective EoS $w_{\rm eff}$ diverges at some high redshift momentarily, then becoming positive and approaching zero at earlier times. This is an unavoidable consequence of the fact that the denominator in Eq.~\eqref{Eq:weff} vanishes at some early time given $m(\theta(a<1)) < m(\theta_0)$ to explain $w_{\rm eff} <-1$ from the DESI data. The critical scale factor $a_c$ when this happens is given parametrically as
\begin{equation}
    a_c = \left[\frac{\rho_{\rm DM}^{(0)}}{\rho_\varphi(a_c)} \left( 1-\frac{m(\theta(a_c))}{m(\theta_0)}\right) \right]^{1/3}\,.
    \label{Eq:ac}
\end{equation}
At scale factor smaller than this critical value, the denominator in Eq.~\eqref{Eq:weff} becomes negative, and hence $w_{\rm eff}$ becomes positive. At even smaller scale factor, the magnitude of the denominator becomes very large due to the $a^{-3}$ scaling, and $w_{\rm eff}$ approaches zero. It is important to emphasize that the apparent singularity in $w_{\rm eff}$ is not physical, but rather an unavoidable consequence of the DM-DE interaction. Consequently, future observational data at high redshifts will be crucial for constraining such interactions. Finally, in the bottom panel, we illustrate the  fit-to-data for our choice of the benchmarks, using DESI+CMB+DESY5~\cite{DESI:2025zgx}, with the horizontal bars indicating the bin width (which is fixed) and the vertical bars representing the 1$\sigma$ error. A qualitatively similar analysis can be done using an exponential form for $V(\varphi)$, however, we find that it becomes more sensitive to the initial conditions.
\subsection{Axion type potential}
\label{sec:axion}
Here we review a technically natural model of interacting DE-DM due to Ref.~\cite{Khoury:2025txd}, where one considers a dark QCD sector associated with a dark axion $\varphi$ playing the role of DE, while the lightest stable dark baryon $\psi$ is identified with the DM. The DM-DE interaction is built into the model as a finite density of $\psi$ modifies the quark condensate, which in turn is connected to the $\varphi$ potential. We UV complete the model and extend it by introducing the anomalous EM coupling as Eq.~\eqref{Eq:LDMDE} to explain the ICB. This model can also help alleviate the $S_8$ tension by delaying the matter-radiation equality to a slightly later time~\cite{Khoury:2025txd}. We will point out an apparent problem of too much DM self-interaction in this setup, and outline technically natural ways to reconcile the model with the observed DM properties.

\begin{table}[htb!]
\centering
\begin{tabular}{|c|c|c|c|c|c|c|} 
\hline 
\rule{0pt}{2.5ex} Fields & $SU(N_{\rm D})$ & $SU(3)_c$ & $SU(2)_L $ & $U(1)_Y$ & $U(1)_{\rm PQ}$\\
\hline\hline
$\Phi$ & 1 & 1 & 1 & 0 & 1 \\
$Q$ & $N_D$ & 1 & 1 & $C_Y$ & 1 \\
$\bar Q$ & $\bar N_D$ & 1 & 1 & $-C_Y$ & 0\\
$\mathbf{u}$ & $N_D$ & 1 & 1 & $0$ & 0\\
$\mathbf{d}$ & $N_D$ & 1 & 1 & $0$ & 0\\
\hline\hline
\end{tabular}
\caption{Field content for the axionic potential model, where $\mathbf{u}$, $\mathbf{d}$ are Dirac fermions, and $Q,\bar Q$ are chiral. 
} 
\label{tab:particles}
\end{table}

Consider an $SU(N_{\rm D})$ dark QCD sector with two light quarks denoted as $\mathbf{u},\mathbf{d}$. A global $U(1)_{\rm PQ}$ Peccei-Quinn (PQ) symmetry is anomalous under $SU(N_{\rm D})$ and is broken spontaneously at a scale $f$, which results in a pseudo-Goldstone boson axion $\varphi$, that transforms as: $\varphi \to \varphi + \mathbb{T} f$ under the $U(1)_{\rm PQ}$ with the transformation parameter $\mathbb{T}$. The field $\varphi$ has the following coupling with dark gluon field strength $G_{\rm D; \mu \nu}^{a}$
\begin{equation}
    {\cal L}_{\rm D} = \frac{g_{\rm D}^2}{32 \pi^2} \frac{\varphi}{f}  G_{\rm D}^{a\,\mu \nu} \widetilde{G}^{a}_{{\rm D};\mu \nu}  \ ,
    \label{Eq:LDGG}
\end{equation}
where $g_{\rm D}$ stands for the $SU(N_{\rm D})$ coupling, and $\widetilde{G}^{a}_{\rm D;\mu \nu} = \frac{1}{2} \epsilon_{\mu \nu \rho \sigma} G_{\rm D}^{a\,\rho \sigma}$. This can be UV completed, for example, by introducing a pair of left-handed chiral fermions $\{Q, \bar Q\}$ having representation $(N_{\rm D},1,1,C_Y)$, $(\bar {N}_{\rm D},1,1,-C_Y)$ under the extended SM gauge symmetry $SU(N_{\rm D}) \otimes SU(3)_c \otimes SU(2)_L \otimes U(1)_Y$, respectively. The bar in $\bar{Q}$ does not denote any charge conjugation. The global PQ charges of $Q, \bar Q$ are $1$, $0$, respectively. A SM singlet scalar $\Phi$ with PQ charge $1$ obtains a vacuum expectation value (VEV) at a scale $f$, spontaneously breaking $U(1)_{\rm PQ}$ and giving masses to $Q, \bar Q$ via its Yukawa coupling $y_{Q}\Phi^* \bar{Q} Q + {\rm h.c.}$ The axion $\varphi$ appears as the axial mode of $\Phi$ after it obtains a VEV. The field content of the model is summarized in Tab.~\ref{tab:particles}. Note that, $Q$, $\bar Q$ can decay to $\mathbf{u,d}$ and SM particles in the early Universe if $C_Y$ is chosen appropriately. For example, the decay can proceed via effective operators like $(\langle \Phi \rangle^*/\Lambda_d^3)   \bar{\mathbf{u}}_R Q e_R e_R$ , $(1/\Lambda_d^2) \bar Q \mathbf{u}_L e_R^c e_R^c$ if $C_Y=2$, while $c$ denotes charge conjugation. Another choice of operators could be $(\langle \Phi \rangle^*/\Lambda_l^3)   \bar{\mathbf{u}}_R Q e_L \nu_L$ , $(1/\Lambda_l^2) \bar Q \mathbf{u}_L e_L^c \nu_L^c$ if $C_Y=1$. As these operators do not violate $U(1)_{\rm PQ}$, they do not contribute to the axion potential. We find that the decay happens much before BBN for any $\Lambda_d < M_{P}$ as $m_Q \sim f$ for order one $y_Q$, thus posing no threat to the standard cosmology. We have also checked that there is no Landau pole for $U(1)_Y$ coupling below $M_{P}$ for $N_{\rm D} C_Y^2 = c_\gamma \lesssim 18$ for $Q, \bar Q$ mass above $10^{12}$ GeV. As $Q, \bar Q$ mass is $\simeq y_Q f$, this is easily satisfied for order one $y_Q$. Note that the light quarks ${\mathbf{q} = \mathbf{u}, \mathbf{d}}$ do not cause any Landau pole as they do not possess any hypercharge.

Once the heavy fermions and the radial mode of $\Phi$ are integrated out, the coupling in Eq.~\eqref{Eq:LDGG} appears due to the given charge assignments of $Q, \bar Q$ and the induced $U(1)_{\rm PQ}-SU(N_{\rm D})^2$ chiral anomaly. At the same time, as $Q, \bar Q$ carry hypercharge, the following coupling with the hypercharge current is generated due to the $U(1)_{\rm PQ}-U(1)_{Y}^2$ anomaly
\begin{equation}
    {\cal L}_Y = (N_{\rm D}\, C_Y^2) \left(\frac{g_{1}^2}{16 \pi^2}\right) \frac{\varphi}{f} B_{\mu \nu} \widetilde{B}^{\mu \nu}\,,
    \label{Eq:Yanomaly}
\end{equation}
where the first factor is the hypercharge anomaly coefficient for the given charge assignment, $g_{1}$ is the $U(1)_Y$ coupling, and $B_{\mu \nu}$, $\widetilde{B}_{\mu \nu}$ are the hypercharge field strength tensor and its dual, respectively. After the electroweak symmetry breaking, Eq.~\eqref{Eq:Yanomaly} generates the EM coupling presented in Eq.~\eqref{Eq:LDMDE}, when $c_\gamma = N_{\rm D}\,C_Y^2$. This is an appealing feature, as for ${\cal O}(1)$ $C_Y$ charges, $c_\gamma \sim {\cal O}(N_{\rm D})$, and hence can be naturally large for large $N_{\rm D}$, and one need not introduce a large number of chiral fermion generations to explain the ICB. Notice that due to the given charge assignment, the domain wall number is $1$, and there is no domain wall problem in the cosmology. Further, all gauge anomaly conditions are satisfied due to vectorial charge assignment for the gauge groups. A high-quality DE axion can be provided by further model building, for example, embedding this in an extra-dimensional setup~\cite{Girmohanta:2023ghm}. 
\begin{table}[htb!]
\centering
\begin{tabular}{|c|c|c|c|c|c|c|} 
\hline 
\rule{0pt}{2.5ex} Benchmarks & $\sigma_\psi/m_\psi$ & $v$ & $\tilde{f}$ & $\Lambda$ [meV] & $\Delta \theta$ & $|c_\gamma|$ 
\\
\hline\hline
BPa & 0.02 & 0  & 0.25  & 2.89 & 0.27 & $11-18$
\\
\hline
BPb & 0.04 & 4 &  0.035 & 1.89 & 3.1 & $1-2$\\
\hline\hline
\end{tabular}
\caption{Benchmark points chosen for Fig.~\ref{Fig:Axion}, for the axion type potential scenario, where we have followed benchmarks of Ref.~\cite{Khoury:2025txd} (that also alleviate the $S_8$ tension). In the last two columns, we have included the resulting field excursion $\Delta \theta$ and corresponding $c_\gamma$, compatible with a 1$\sigma$ value of the ICB angle $\beta= (0.261 \pm 0.061)^\circ$~\cite{Lin:2025gne}. We have chosen $m_{\mathbf{u}} = 0.8 m_{\mathbf{d}}$, and the initial value of $\theta=2.5,2$ for BPa, BPb, respectively.}  
\label{tab:BPAxion}
\end{table}

\begin{figure}[htb!]
    \centering    
    \includegraphics[scale=0.3]{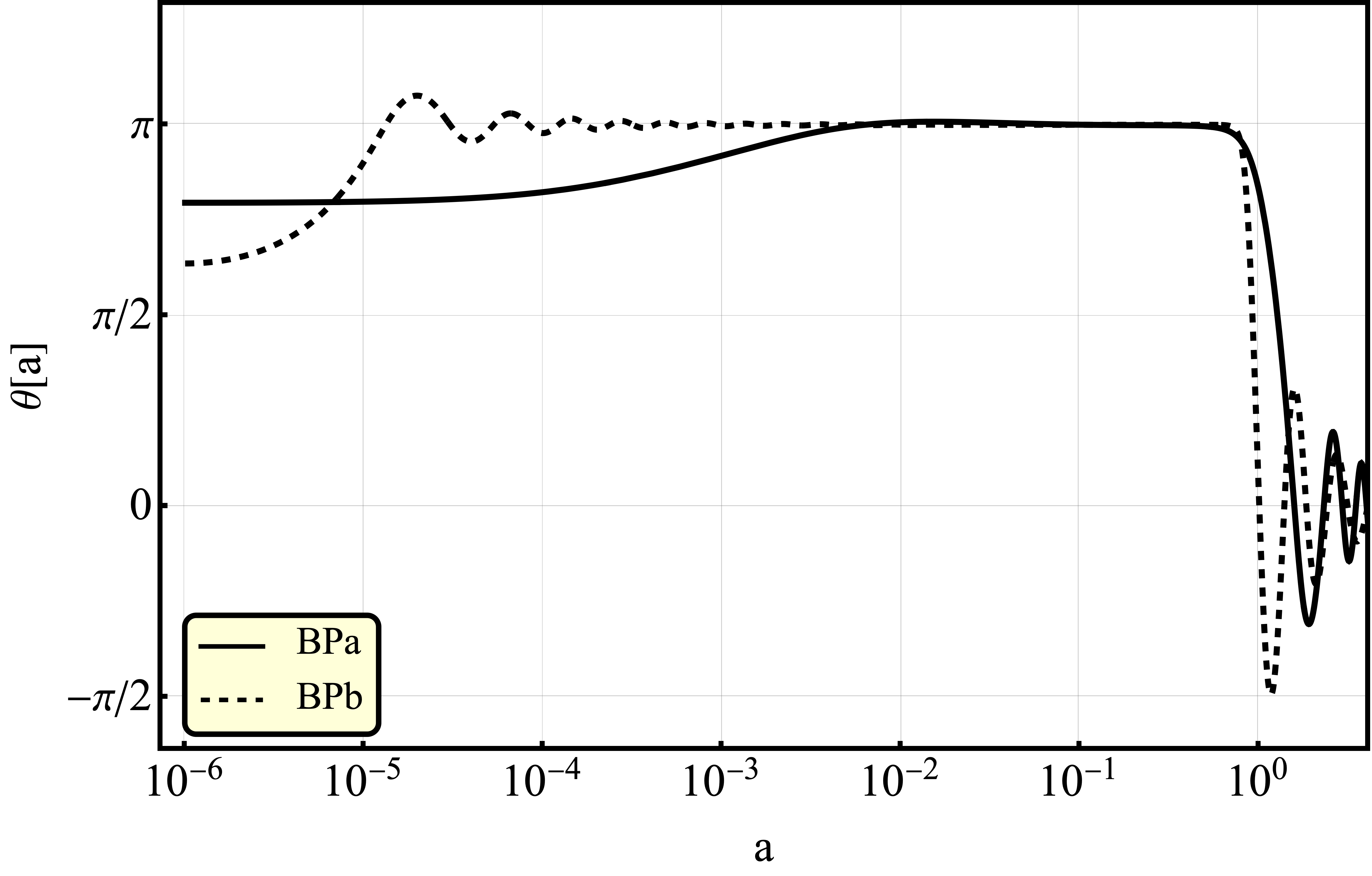}~
    \includegraphics[scale=0.32]{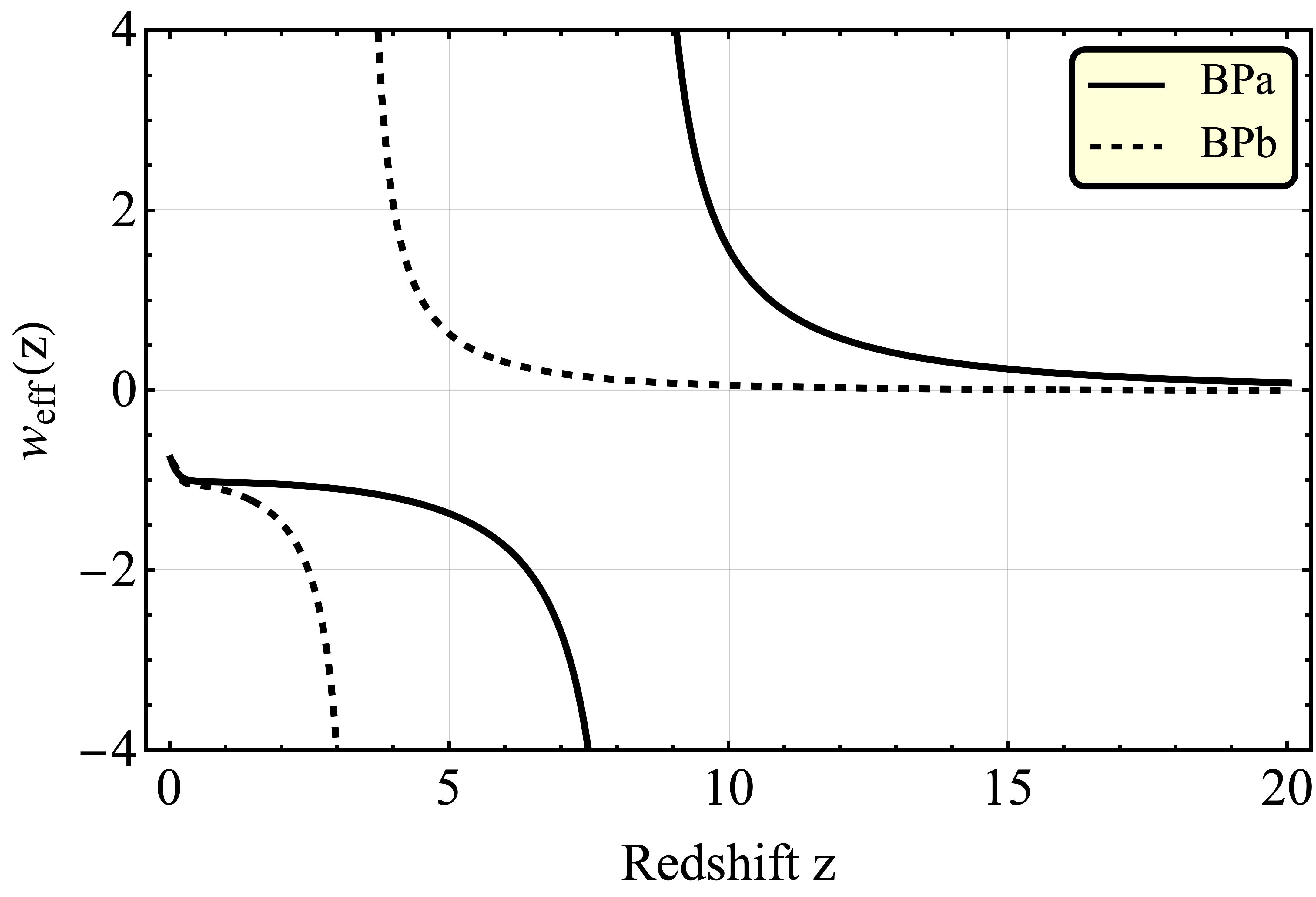}\\[10pt]
    \includegraphics[scale=0.4]{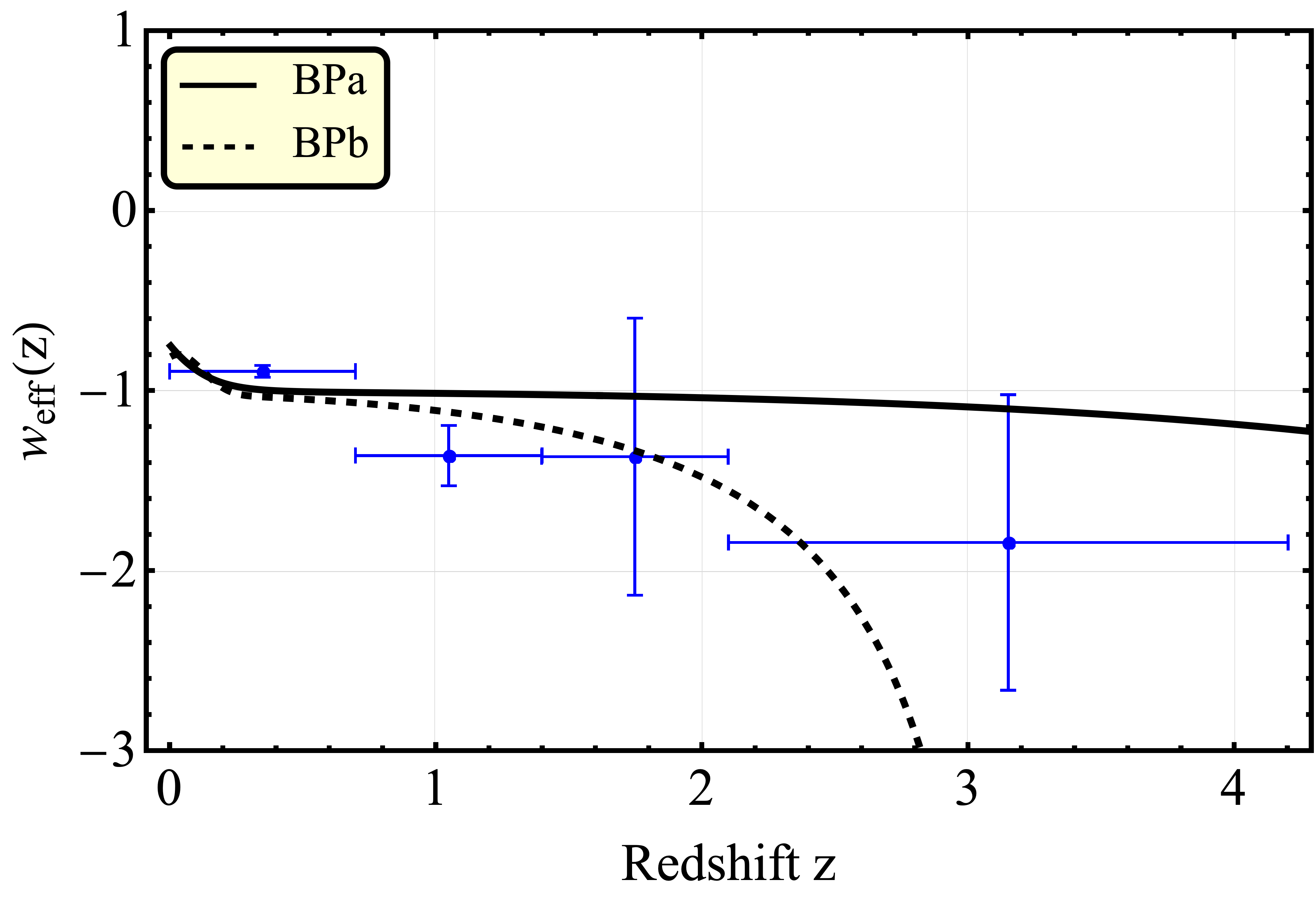}
    \caption{Similar to Fig.~\ref{fig:fitPoly} for the axion potential in Eq.~\eqref{Eq:VeffAxion} following Ref.~\cite{Khoury:2025txd}. All relevant parameters are mentioned in the inset and tabulated in Tab.~\ref{tab:BPAxion}.}
    \label{Fig:Axion}
\end{figure}

Once the dark QCD confines, it generates a potential for $\varphi$ due to Eq.~\eqref{Eq:LDGG} as follows
\begin{equation}
    V_{\rm vac}(\theta) = \Lambda^4 \left[1+v-\sqrt{1-\Upsilon \sin^2\left(\frac{\theta}{2} \right)} \right] \ ,
    \label{Eq:V0Axion}
\end{equation}
where $\theta = \varphi/f$, $v$ is a constant shift of the potential, which is allowed in general, and $\Upsilon = 4 m_{\mathbf{u}} m_{\mathbf{d}}/(m_{\mathbf{u}}+m_{\mathbf{d}})^2$, while $m_{\mathbf{u}}, m_{\mathbf{d}}$ denote the light dark quark masses. The field is periodic $\theta=\theta+2 \pi$ and this is reflected in the potential.  The scale $\Lambda$, which sets the scale of the DE potential in Eq.~\eqref{Eq:V0Axion} is related to the quark condensate 
\begin{equation}
    \Lambda^4 = -\varepsilon (m_{\mathbf{u}}+m_{\mathbf{d}}) \langle \bar {\mathbf q}  {\mathbf q} \rangle_{\rm vac}  \ ,
    \label{Eq:LambdaDE}
\end{equation}
where a model-dependent parameter $\varepsilon < 1$ is assumed to have control over the finite density correction, similar to light axion models~\cite{Hook:2017psm, Hook:2018jle}, and $\langle \bar{\mathbf{u}}  \mathbf{u} \rangle \simeq \langle \bar{\mathbf{d}}  \mathbf{d} \rangle$ is assumed, while $\mathbf{q}$ represents $\mathbf{u},\mathbf{d}$. The DM-DE interaction is naturally built into this model as Eq.~\eqref{Eq:V0Axion} is only valid in the vacuum, and in the background of finite density of the dark baryon DM $\psi$, the quark condensate is modified~\cite{Cohen:1991nk, Hook:2017psm, Balkin:2020dsr}, namely
\begin{align}
    \langle \bar {\mathbf q}  {\mathbf q} \rangle_{\psi} = \langle \bar {\mathbf q}  {\mathbf q} \rangle_{\rm vac} + \frac{\partial \Delta E}{\partial m_{\mathbf q}} \ ,
    \label{Eq:qbarq}
\end{align}
where $\Delta E$ is the dark QCD ground state energy shift at finite density of $\psi$. Treating the DM as a free Fermi gas, \textit{i.e.,} neglecting its interactions, in the non-relativistic limit, $\Delta E \simeq m_\psi n_\psi$, where $m_\psi$ is the DM mass and $n_\psi$  denotes its density. Therefore, the effective DE potential at finite $n_\psi$ is~\cite{Hook:2017psm, Khoury:2025txd}
\begin{align}
    V_{\rm eff}(\varphi) & = \Lambda^4 \left[ 1+ v - \left( 1-\frac{ n_\psi \sigma_\psi}{\Lambda^4} \right) \sqrt{1-\Upsilon \sin^2\left(\frac{\theta}{2} \right)}  \right] + {\cal O} \left(\varepsilon^2 \frac{n_\psi^2 \sigma_\psi^2}{\Lambda^8} \right) \ ,
    \label{Eq:VeffAxion}
\end{align}
where $m_{\mathbf{u}} \simeq m_{\mathbf{d}}$ is assumed, and
\begin{equation}
    \sigma_\psi \equiv \sum_{{\mathbf q}=\mathbf{u},\mathbf{d}} \frac{\partial m_\psi}{\partial \ln m_{\mathbf q}} \ .
    \label{Eq:sigmaDef}
\end{equation}
Importantly, $\sigma_\psi > 0$, as heavier constituent quarks imply heavier baryons. Hence, the effective potential is minimized at $\theta=\pi$ for DM density above the critical number density $n^{c}_\psi = \Lambda^4/\sigma_\psi$, while below this density the minimum switches to $\theta=0$. For $\varepsilon<1$, perturbative control is maintained at the critical density. Interestingly, this critical density scale and associated dynamics are relevant for the late-time Universe for DE axion, while for QCD axion, the finite density effect is only relevant inside a neutron star~\cite{Hook:2017psm}. From Eq.~\eqref{Eq:VeffAxion} one can identify
\begin{equation}
    \frac{m(\theta)}{m_\psi} = 1 + \frac{\sigma_\psi}{m_\psi}  \sqrt{1-\Upsilon \sin^2\left(\frac{\theta}{2} \right)} \ ,
    \label{Eq:mthetaAxion}
\end{equation}
and $\Upsilon \simeq 1$ for $m_{\mathbf{u}} \simeq m_{\mathbf{d}}$ such that for present-day values of $\theta \to 0$, the DM mass is $m_{\rm DM} \simeq m_\psi$, as $\sigma_\psi/m_\psi \simeq {\cal O}(10^{-2})$ as shown in Tab.~\ref{tab:BPAxion}.

In Fig.~\ref{Fig:Axion} we demonstrate the evolution of the axion field and the resulting equation of motion following Ref.~\cite{Khoury:2025txd}. The field naturally finds itself at $\theta=\pi$ irrespective of the initial condition, and evolves to $\theta=0$ in the late Universe, once the DM density has dropped below $n_\psi^{c}$. This is a very appealing feature, as it naturally explains why the axion finds itself near the hilltop of the late-time effective potential to explain the DE today. In conventional quintessence axion models, this initial condition is chosen by hand. We obtain the resulting field excursion $\Delta \theta$ from the solution to the EoM for axion, and evaluate the required $c_\gamma$ to explain the ICB reported, as shown in Tab.~\ref{tab:BPAxion}. As $c_\gamma = N_{\rm D}\,C_Y^2$, it is relatively easier to accommodate the required charge with ${\cal O}(1)$ $C_Y$ for an appropriate number of dark colors $N_{\rm D}$. To be quantitative, for the BPa, one can choose $N_{\rm D}=3$, and $C_Y=2$, while for BPb, $N_{\rm D}=2$, $C_Y=1$ is sufficient to obtain the required $c_\gamma$ as shown in Tab.~\ref{tab:BPAxion}. The chosen benchmark can fit the DESI results, while when extrapolated to higher redshift, the EoS approaches $0$ as consistent with the generic interacting DE-DM framework outlined in Sec.~\ref{sec:framework}.

Before moving on, let us highlight a few important facts here. To explain the DE energy density today, $\Lambda \simeq {\cal O}(\rm meV)$, as shown in Tab.~\ref{tab:BPAxion}. Further, the dark QCD scale where confinement sets in is given by $\Lambda_{\rm dQCD}^3 \simeq -\langle \bar {\mathbf q} {\mathbf q} \rangle_{\rm vac}$. Only when the DS temperature, which in principle can be different from the SM temperature bath in the early Universe, is below $\Lambda_{\rm dQCD}$, the DM is formed and the axion potential is generated. From Eq.~\eqref{Eq:LambdaDE}, we arrive at the following relation
\begin{equation}
    \Lambda_{\rm dQCD} \simeq \Lambda \left( \frac{\Lambda}{2 \varepsilon \bar{m}_{\mathbf q}} \right)^{1/3} \ ,
    \label{Eq:LambdadQCDrel}
\end{equation}
where the average light quark mass $\bar{m}_{\mathbf q} = (m_{\mathbf{u}}+m_{\mathbf{d}})/2$.  Note that $m_{\mathbf{q}} \neq 0$, \textit{i.e.,} no dark quark is massless, otherwise $\varphi$ is rendered unphysical and can be removed by chiral rotation of the massless dark quark. If there is no large hierarchy between $\Lambda$ and $\varepsilon \bar{m}_{\mathbf q}$, then $\Lambda_{\rm dQCD} \simeq {\cal O}(\rm meV)$. This is problematic as the DM $\psi$ is a light baryon of the dark QCD; it naturally inherits self-interaction via the mediation of dark pions. Further, $\Lambda_{\rm dQCD}$ sets all the scales for the DM, including the DM mass and the scattering length. Therefore, in the absence of some screening mechanism due to additional dark force carriers, the DM self-interaction cross-section per unit DM mass can be estimated just from dimensional analysis and is expected to be
\begin{equation}
    \frac{\sigma_{\rm DM-DM}}{m_\psi} \sim {\cal O}\left(\frac{1}{\Lambda_{\rm dQCD}^3} \right) \ .
    \label{Eq:sigmaDMDM}
\end{equation}
If $\Lambda_{\rm dQCD} \sim {\cal O}$(meV), then this self-interaction cross-section is extremely large, and is at odds with the collisionless property of the DM observed in the dynamics of galactic and cluster scales~\cite{Tulin:2017ara, Girmohanta:2022dog, Girmohanta:2022izb}. However, if $\varepsilon \bar{m}_{\mathbf{q}} \ll \Lambda$, then $\Lambda_{\rm dQCD} \gg \Lambda$, and this problem can be solved in a technically natural way.\footnote{We do not worry about naturalness issues related to tiny Dirac constituent quark masses, as it is well-known that additional structures, such as embedding in an extra-dimensional framework, may render the masses tiny due to the corresponding wavefunction overlaps~\cite{Grossman:1999ra,Girmohanta:2021gpf}.} To be more quantitative, we can estimate the DM self-interaction strength per unit DM mass as follows~\cite{Kribs:2016cew}
\begin{align}
    \frac{\sigma_{\rm DM-DM}}{m_\psi} \simeq 1 { \ \rm cm}^2/{\rm g} \ \left( \frac{\Lambda_{\rm dQCD}}{m_{_\psi}}\right) \left( \frac{\Lambda_{\rm dQCD}}{a_{\rm D}^{-1}}\right)^2 \left( \frac{0.15 { \ \rm GeV}}{\Lambda_{\rm dQCD}}\right)^3 \ ,
    \label{Eq:SIDM}
\end{align}
where $a_{\rm D}$ denotes the effective scattering length, and $1 { \ \rm cm}^2/{\rm g}$ is the relevant cross-section for diversity of galactic rotation curves~\cite{Tulin:2017ara}. Therefore, if $\Lambda_{\rm dQCD} \gtrsim 0.1$ GeV, then the too much self-interaction problem can be solved, and if $\Lambda_{\rm dQCD}$ lies in the range relevant for small-scale structure issues, it can be a virtue of the model. Furthermore, a first-order phase transition in the dark QCD sector can explain the nano-Hz stochastic gravitational wave signal if $\Lambda_{\rm dQCD} \sim 0.1-1$ GeV~\cite{Fujikura:2023lkn}. The DM mass can be estimated by scaling up the QCD formula~\cite{Chivukula:1989qb}
\begin{align}
    m_\psi \simeq m_p \left(\frac{\Lambda_{\rm dQCD}}{\Lambda_{\rm QCD}} \right) \left(\frac{N_{\rm D}}{3}\right) \ ,
    \label{Eq:DMMassAxion}
\end{align}
where $m_p$, $\Lambda_{\rm QCD}$ denote the proton mass and QCD dynamical scale, respectively. We outline two possibilities to accommodate the DM relic density in such a scenario.
\section{Dark matter in the light of DESI \& ICB}
\label{sec:dm}
In this section, we examine how the DM obtains its relic density within the minimal framework defined by the interaction Lagrangian in Eq.~\eqref{Eq:LDMDE}. By \textit{minimal} we imply that the DM relic density is obtained by only using the interactions available in the aforementioned models, \textit{i.e.,} without introducing any additional fields or couplings. Interestingly, the form of the DE potential gives rise to two qualitatively distinct mechanisms for DM production. In the case of a polynomial potential, the DM can be produced via a freeze-in mechanism utilizing the dimension-5 EM interaction that explains the observed ICB. The DM generated via this process is typically heavy and is produced non-thermally via 2-to-2 annihilation of particles in the thermal bath, with the production rate dictated by the scale $f$. On the other hand, the axion-like potential points to a sub-GeV DM candidate, whose relic abundance can arise from thermal freeze-out processes occurring entirely within the DS, or via sharing of primordial particle number asymmetry via minimal field extension, as elaborated below.
\subsection{Dark matter in polynomial potential}
The most straightforward way to produce DM following the interaction Lagrangian in Eq.~\eqref{Eq:LDMDE}, is from the thermal bath via $\gamma\gamma\to\psi\psi$, mediated by $\varphi$ in the $s$-channel. As the relevant energy scale for this freeze-in DM production is the reheat temperature, we use the running coupling appropriately. We will assume that the full SM gauge invariant Lagrangian, relevant at temperatures where the electroweak symmetry is restored is given by
\begin{equation}
    {\cal L}_{\rm ICB} = c_\gamma \left(\frac{\alpha_1}{4 \pi}\right) \frac{\varphi}{f} B_{\mu \nu} \widetilde{B}^{\mu \nu} \ ,
    \label{Eq:ICBtrh}
\end{equation}
where the $U(1)_Y$ field strength tensor and its dual are denoted by $B_{\mu \nu}$, $\widetilde{B}_{\mu \nu}$, respectively, and its gauge coupling strength is $\alpha_1 = g_1^2/(4 \pi)$. The scattering cross-section for $\gamma\gamma\to\psi\psi$ mediated by $\varphi$ is given as follows
\begin{align}
\sigma(s)_{\gamma\gamma\to\psi\psi}=\frac{c_\gamma^2\,\kappa^2\,\alpha_1^2}{64\,\pi^3}\,\frac{\mdm^2\,s^2}{f^4\,\left[(s-m_\varphi^2)^2+\Gamma_\varphi^2\,m_\varphi^2\right]}\,\left(1-\frac{4\,\mdm^2}{s}\right)^{3/2}\,,    
\end{align}
which, in the limit $s\gg m^2$ becomes,
\begin{align}
&\sigma_{\gamma\gamma\to \psi\psi}\simeq \frac{c_\gamma^2 \kappa^2 \alpha_1^2 m_\psi^2}{\left(4\pi\right)^3\,f^4 }\,. 
\end{align}
Note that the DM mass is given by Eq.~\eqref{Eq:poly}. In order to track the evolution of DM number density with time, we write down the Boltzmann equation (BEQ) as follows,
\begin{align}
\dot n_{\rm DM}+3\,H\,n_{\rm DM}=\Gamma\,,    
\end{align}
where the reaction density in case of 2-to-2 scattering is given by~\cite{Edsjo:1997bg,Gondolo:1990dk,Duch:2017khv},
\begin{align}
& \Gamma=\frac{T}{8\,\pi^4}\,\int_{4\mdm^2}^\infty ds\,s^{3/2}\,\sigma\left(s\right)_{\gamma\gamma\to \psi\psi}\,K_1\left(\frac{\sqrt{s}}{T}\right)\label{eq:gam-ann}\,,
\end{align}    
while $K_1$ denotes the modified Bessel function of the 1st-order. In a radiation-dominated Universe, where the entropy per comoving volume is conserved,  the above equation can be recast in terms of the DM yield $Y_{\rm DM}=n_{\rm DM}/s$ as
\begin{equation}\label{eq:beq}
x\,H\,\mathfrak{s}\,\frac{dY_{\rm DM}}{dx} =\Gamma(T)\,,
\end{equation}
where $x\equiv\mdm/T$ is a dimensionless quantity with $T$ being the temperature of the thermal bath. For a radiation-dominated Universe, the entropy density $\mathfrak s$ and Hubble parameter $H$ are given by, 
\begin{align}
& \mathfrak{s}(T)=\frac{2\,\pi^2}{45}\,\gss(T)\,T^3\,, & 
H(T)=\frac{\pi}{3}\,\sqrt{\frac{\gs(T)}{10}}\,\frac{T^2}{M_P}\,,
\end{align}
where $\gss$ and $\gs$ are the effective number of relativistic degrees of freedom contributing to the entropy and energy density, respectively. 
\begin{figure}[htb!]
    \centering    
    \includegraphics[scale=0.375]{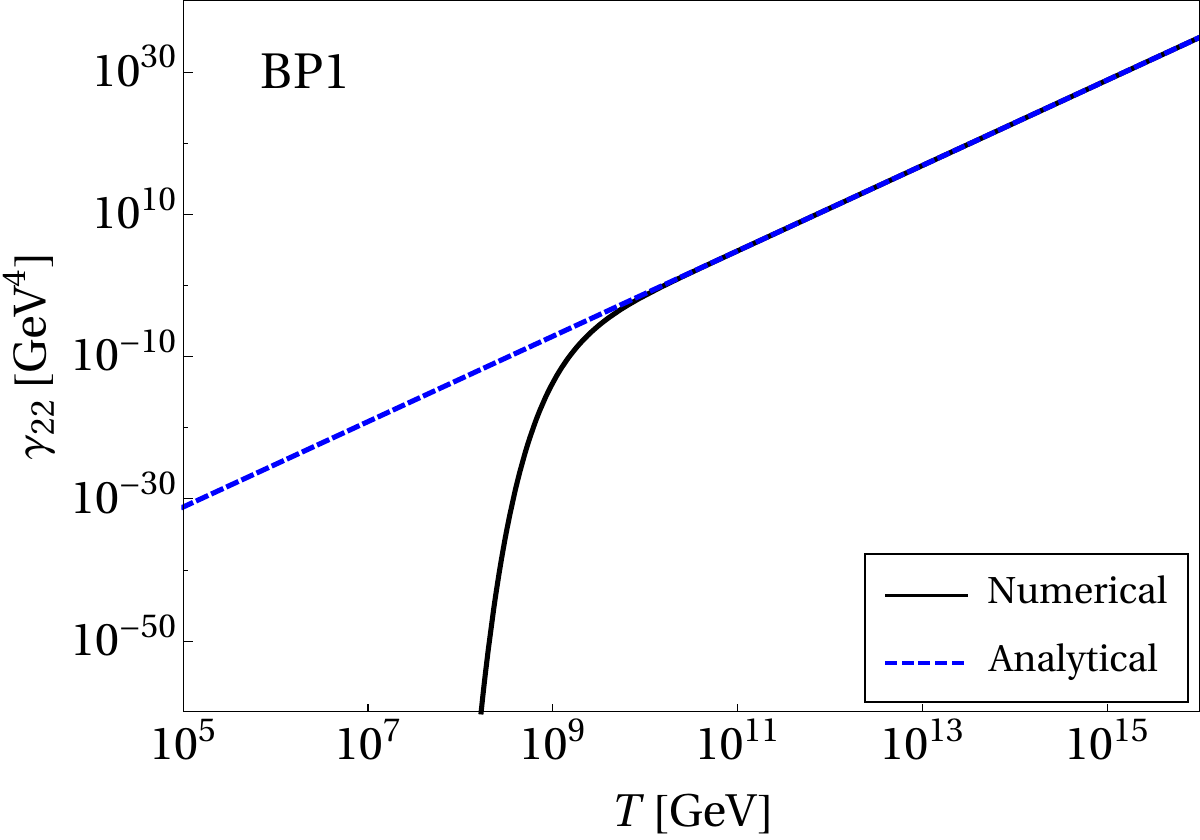}~
    \includegraphics[scale=0.275]{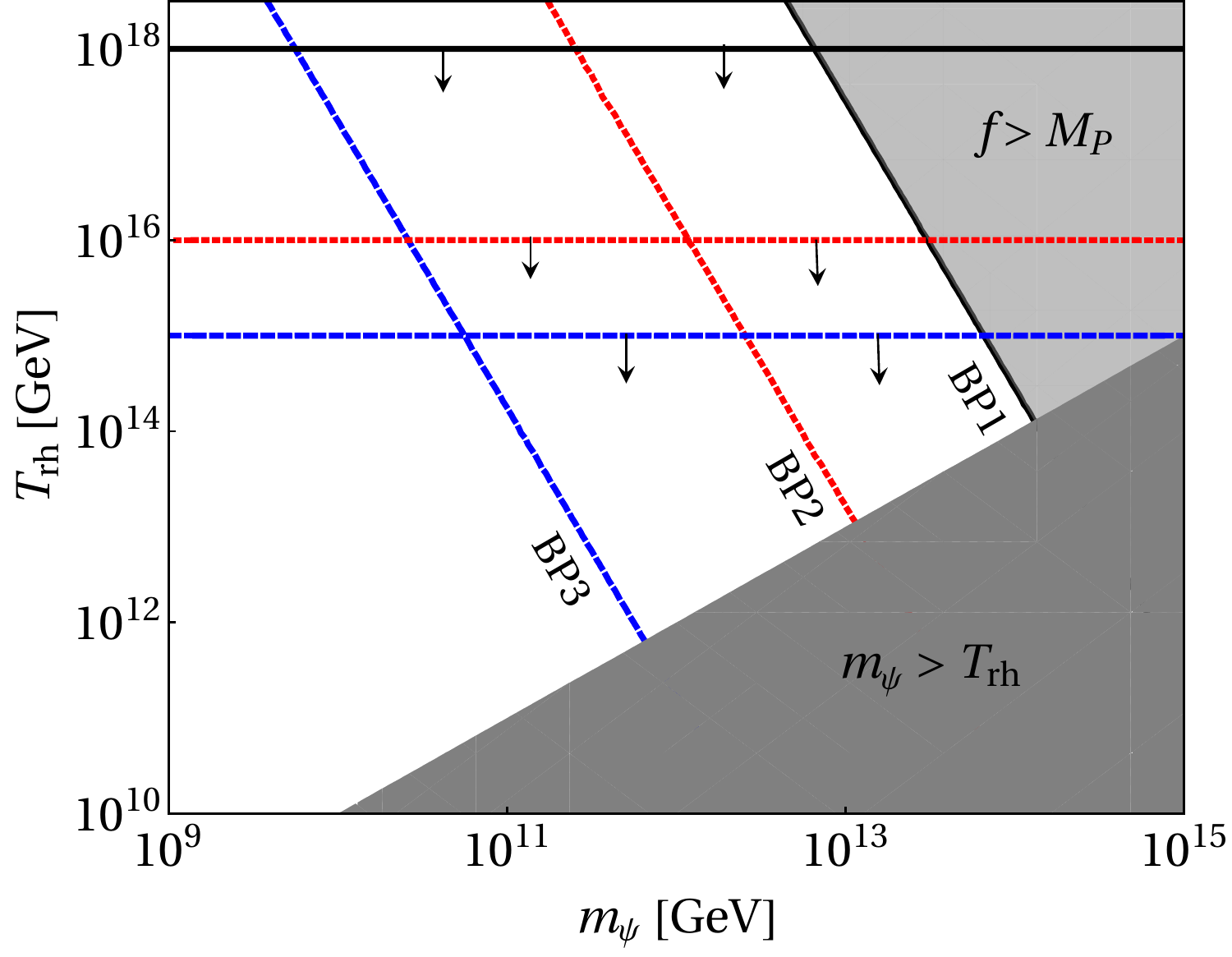}\\[10pt]
    \includegraphics[scale=0.45]{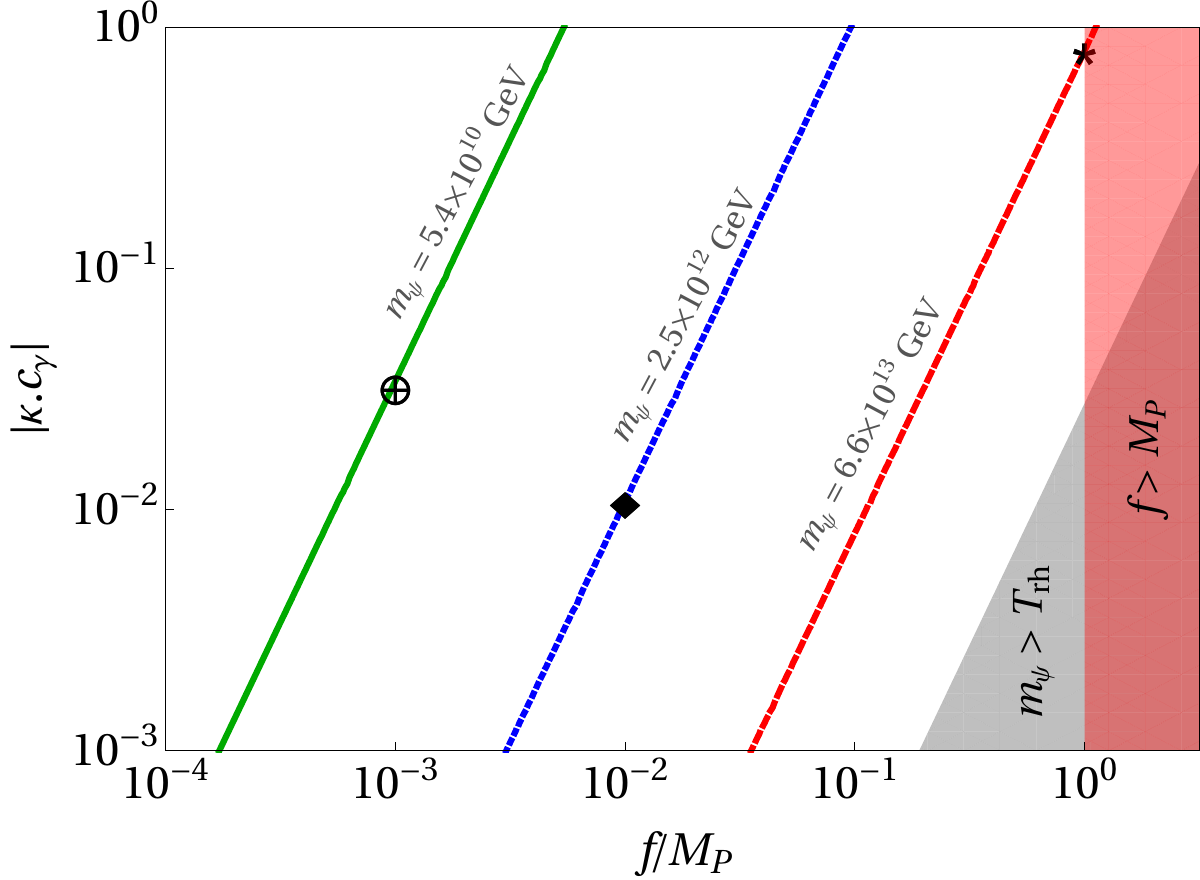}
    \caption{{\it Top left}: Comparison of 2-to-2 reaction density, obtained analytically (blue dashed) with the one obtained numerically (black solid), for BP1, considering a DM of mass $10^{10}$ GeV. {\it Top right}: Contours of right relic abundance in $\Trh-\mdm$ plane, corresponding to the benchmarks in Tab.~\ref{tab:BP}. The arrowheads indicate the viable region of the parameter space that satisfies the EFT bound $f>\Trh$ for each of the benchmarks. {\it Bottom:} Contours of right DM abundance for a fixed $\Trh=10^{15}$ GeV, where different contours correspond to different DM masses. The benchmarks BP1, BP2, and BP3 in Tab.~\ref{tab:BP} are denoted by the star, box, and crossed-circle, respectively. The shaded regions are disallowed from the instantaneous reheating condition $(\mdm<\Trh)$ and super-Planckian value of the effective scale $(f<M_P)$.}
    \label{fig:freezein}
\end{figure}
In the limit $s\gg m^2$, the reaction density takes the form
\begin{align}\label{eq:gam22}
& \Gamma\simeq \frac{c_\gamma^2\,\kappa ^2\,\alpha_1^2}{16\,\pi^7}\,\frac{T^6\,\mdm^2}{f^4}\,.     
\end{align}
Note that the reaction density shows a strong dependence on temperature, which is a typical feature of
UV freeze-in~\cite{Hall:2009bx,Elahi:2014fsa}. The left panel of Fig.~\ref{fig:freezein} shows a comparison between analytically and numerically computed reaction densities. For $T\gtrsim\mdm$, they exactly match with each other, while for $T\lesssim\mdm$, we see a difference as the bath temperature is not enough to produce a pair of DM, which is taken into account numerically.  

Using Eq.~\eqref{eq:gam22}, one can now solve the BEQ in Eq.~\eqref{eq:beq} analytically and obtain the asymptotic DM yield as
\begin{align}\label{eq:Ydm}
& Y_{\rm DM}(T_0\ll\Trh)\simeq\frac{27\,c_\gamma^2\,\kappa^2\,\alpha_1^2}{640\,\pi^{10}\,\sqrt{10}}\,\frac{M_P\,\Trh}{f^2}\,\left(\frac{\mdm}{f}\right)^2\,,
\end{align}
where it is assumed that $\gs,\gss \simeq 100$, and $Y_{\rm DM}(\Trh)\simeq 0$. This is the crucial assumption of freeze-in, where the DM number density is close to zero at the highest temperature of the Universe, and with time, it is produced from the visible sector~\cite{Hall:2009bx,Bernal:2017kxu,Elahi:2014fsa}. To saturate the observed DM abundance, it is required to have
\begin{equation} \label{eq:obsyield}
    Y_0\, \mdm = \Omega_{\rm DM}^{(0)} h^2 \, \frac{1}{s_0}\,\frac{\rho_c}{h^2} \simeq 4.3 \times 10^{-10}\,\text{GeV}\,,
\end{equation}
where $Y_0 \equiv Y_{\rm DM}(T_0)$ is the DM yield today. Furthermore, $\rho_c \simeq 1.05 \times 10^{-5}\, h^2$~GeV/cm$^3$ is the critical energy density, $s_0\simeq 2.69 \times 10^3$~cm$^{-3}$ the present entropy density~\cite{ParticleDataGroup:2022pth}, and $\Omega_{\rm DM}^{(0)} h^2 \simeq 0.12$ the observed abundance of DM relics~\cite{Planck:2018vyg}. From the above relation, we find,
\begin{align}
& \Trh\simeq 8\times 10^{14}\,\text{GeV}\left(\frac{7\times10^{13}\,\text{GeV}}{\mdm}\right)^3\,\left(\frac{f}{M_P}\right)^4\,\left(\frac{8}{\kappa\cdot c_\gamma}\right)^2\,,
\end{align}
for BP1, where we have assumed $\gs\approx\gss\simeq 100$. Clearly, for a fixed DM mass, a larger $f$ requires a higher reheating temperature since that results in DM under abundance. For the previous analysis to be valid, the DM has to be out of chemical equilibrium with the SM bath. One needs to guarantee, therefore, that the interaction rate density be $\Gamma< n_{\rm eq}^{\rm DM}\,H$ at $T=\Trh$. This results in an upper bound on the reheating temperature,
\begin{align}\label{eq:Trh-bound}
\Trh <  1.6\times 10^{14}\,\text{GeV}\,\left(\frac{f}{2\times 10^{-6}\,M_P}\right)^4\,\left(\frac{10^{12}\,\text{GeV}}{\mdm}\right)^2\,\left(\frac{10}{\kappa\cdot c_\gamma}\right)^2\,,
\end{align}
where we consider $n_{\rm eq}^{\rm DM}(T)=\frac{T}{2\pi^2}\,\mdm^2\,K_1\left(\mdm/T\right)$ to be the equilibrium DM number density, such that it provides a conservative limit on $\Trh$. For our choice of parameters, Eq.~\eqref{eq:Trh-bound} is always satisfied. We emphasize that, for the benchmarks shown in Table~\ref{tab:BP}, the DM--DE interaction modifies the DM mass at an earlier cosmological time at the level of $\sim 3\%$ for BP1, while its $\lesssim 1\%$ for the other benchmarks. To keep the final abundance unchanged, this variation can be reabsorbed in the unknown reheat temperature according to Eq.~\eqref{eq:obsyield}. 

The top right panel of Fig.~\ref{fig:freezein} illustrates the allowed parameter space in the $\Trh$--$\mdm$ plane that yields the correct relic abundance, with different contours corresponding to the benchmark points listed in Tab.~\ref{tab:BP}. As $\tilde f$ decreases from BP1 to BP3 (alongside a reduction in $|\kappa \cdot c_\gamma|$ ), the contours consistent with the observed DM relic density shift toward lower values of both $\Trh$ and $\mdm$. This behavior aligns with expectations from Eq.~\eqref{eq:Ydm}, ensuring that the Universe is not overclosed. Assuming instantaneous reheating, the thermal bath cannot exceed a temperature $\Trh$, which restricts the DM mass to be below $\Trh$; hence, the shaded region at the bottom right is excluded. On the other hand, super-Planckian values of $f$ are disallowed, excluding the top right shaded region. Additionally, to maintain the validity of the effective field theory (EFT), $f < \Trh$ is not permitted. As a result, the arrowheads for each benchmark indicate the allowed parameter space. Among the benchmarks, BP1 provides the largest viable region (since $f=M_P$) that satisfies the DM relic abundance, is consistent with DESI observations, and accommodates the measured isotropic cosmic birefringence. 
The bottom panel of Fig.~\ref{fig:freezein} shows contours of correct relic abundance for a fixed reheating temperature $\Trh = 10^{15}$ GeV, across varying DM masses. As expected from Eq.~\eqref{eq:Ydm}, a larger $\mdm$ demands a higher $f$ to avoid overproduction. The benchmark points listed in Tab.~\ref{tab:BP} are shown here as well, where $c_\gamma$ is evaluated at the central value of $\beta=0.261$. Altogether, the combined parameter space is consistent within $1\sigma$ of DESI+CMB+DESY5 data and the ICB angle, pointing toward a scenario with high reheating temperatures and heavy DM. 

Finally, two comments in order: (i) precision measurements of primordial element abundances from Big Bang nucleosynthesis suggest a lower bound on the reheating temperature of $\Trh \gtrsim \text{few MeV}$~\cite{Sarkar:1995dd, Kawasaki:2000en,Hannestad:2004px, DeBernardis:2008zz, deSalas:2015glj,Hasegawa:2019jsa}. On the other hand, models of inflation typically imply an upper bound around $\Trh \lesssim 10^{16}\,\mathrm{GeV}$ (see, e.g., Ref.~\cite{Linde:1990flp}). Moreover, a high reheating temperature can, in principle, lead to the overproduction of long-lived exotic relics, which may overclose the Universe. A classic example of such a problem is the cosmological gravitino problem in supergravity theories~\cite{Moroi:1993mb}. This consideration typically leads to an upper bound on the reheating temperature in supergravity models of around $\Trh \lesssim 10^{10}\,\mathrm{GeV}$, and (ii) beyond the instantaneous reheating approximation, the temperature of the thermal bath can reach a maximum value $T_{\mathrm{max}}$ that exceeds the reheating temperature $\Trh$~\cite{Giudice:2000ex}. It is conceivable that the DM abundance is established during this reheating phase, in which case its final abundance can differ substantially from freeze-in estimates based on the assumption of radiation domination, as shown, for example, in~\cite{Garcia:2017tuj,Bernal:2019mhf,Chen:2017kvz}. However, a precise determination of $\Tmax$ and $\Trh$ depends on the underlying model of inflation and the post-inflationary inflaton dynamics, which is beyond the scope of the present study. 
\subsection{Dark matter in axion-type potential}
As discussed in the previous subsection, the freeze-in production mechanism for $\psi$ requires both a very high reheating temperature ($\Trh \gtrsim 10^{15}~\text{GeV}$) and an extremely heavy dark matter particle ($m_{\rm DM} \gtrsim 10^{10}~\text{GeV}$), with both scales sensitive to the choice of $f$ required to fit the DESI data. In contrast, for the model presented in subsection~\ref{sec:axion}, the DM mass is determined by the dark QCD dynamical scale $\Lambda_{\rm dQCD}$ as shown in Eq.~\eqref{Eq:DMMassAxion}. As discussed in the subsection~\ref{sec:axion}, to be consistent with the collisionless property of DM at large scales, $\Lambda_{\rm dQCD} \gtrsim 0.1$ GeV, and can be achieved by choosing the constituent quark mass appropriately. Moreover, as shown in Tab.~\ref{tab:BPAxion}, the decay constant $f$ must remain sub-Planckian to be consistent with DESI data, while neglecting quantum gravity effects. Taken together, these considerations render the freeze-in mechanism incompatible with this scenario. Instead, we identify two promising alternatives to achieve the observed DM relic abundance. Since these mechanisms are well-established in the literature, we refrain from providing detailed calculations here and instead present a more qualitative overview below.
\begin{itemize}
\item [(i)] {\it {\underline{Dark freeze-out}}:} The dark baryons $\psi$ can undergo freeze-out within the DS via, for example, $\bar\psi\psi\to\pi_{\rm D}\,\pi_{\rm D}$, where $\pi_{\rm D}$ are the dark pions. The dark quarks $Q, \bar Q, {\mathbf{u,d}}$ can remain in equilibrium with each other via the mediation of dark gluons. Therefore, a notion of dark temperature $T_{\rm D}$ can be attributed. Despite $Q, \bar Q$ carrying hypercharge quantum numbers, and in the presence of dimension-5 EM coupling and effective operators relevant for $Q, \bar Q$ decay as mentioned in subsection~\ref{sec:axion}, it is easy to ensure that the DS never comes in thermal equilibrium with the visible sector in the entire thermal cosmological history due to very large $f$. As long as one satisfies $T_{\rm rh} < f$, the heavy dark quarks $Q, \bar Q$ never become massless in the cosmological evolution. For example, consider the scattering $\gamma Q \to \gamma Q$ in the early Universe. It is sufficient to satisfy that the scattering rate for this process is less than the Hubble rate at the reheating temperature. This is satisfied when
\begin{align}
    T_{\rm rh} & < M_P \left( \frac{1.4}{C_Y}\right)^4 \left( \frac{y_Q\,f}{\alpha_1 M_P} \right)^2\,,
    \label{Eq:rhcon2}
\end{align}
obtained by comparing the interaction rate $n_{\rm eq}\,\langle\sigma\cdot v\rangle$ with the Hubble rate in a radiation dominated background, where we consider $\langle\sigma\cdot v\rangle\simeq C_Y^4\,\pi\,\alpha_1^2/m_Q^2$ and $n_{\rm eq}=\left(2\,\zeta(3)/\pi^2\right)\,T^3$ is the equilibrium number density of photons. The condition in Eq.~\eqref{Eq:rhcon2} is easily satisfied with the chosen benchmark parameters in subsection~\ref{sec:axion}. Similarly, we have checked that it is straightforward to satisfy the out-of-equilibrium condition for the dimension-5 EM coupling and the effective operators relevant for $Q, \bar Q$ decay. Furthermore, if one considers the operator $\langle \Phi \rangle^*/\Lambda_d^3 \bar{\mathbf{u}}_R Q e_R e_R$, then the DS never comes in thermal equilibrium with the SM, provided
\begin{align}
    \Lambda_d >  10^{15} {\,\rm GeV} \left( \frac{y_Q}{10^{-3}} \right)^{1/3} \left( \frac{T_{\rm rh}}{10^{15} { \,\rm GeV}} \right)^{1/6} \left( \frac{f}{10^{-2} M_P} \right)^{2/3} \ ,
    \label{Eq:lambdaD}
\end{align}
while $Q, \bar Q$ decays to the light dark quarks and electrons much before BBN via this operator. Therefore, we can consistently assume that the DS temperature $T_{\rm D}$ is distinct from the visible sector bath, which may have its origin in how the inflaton couples to different fields and the reheating history after inflation.

As the temperature $T_{\rm D}$ of the hidden sector falls below $m_\psi$, the number density of $\psi$ particles $n_\psi$ tracks the equilibrium distribution and becomes exponentially suppressed. Ultimately, $\psi$ undergoes freeze-out within the DS once $n_\psi\,\langle\sigma\cdot v\rangle=H$ at $T_{\rm D}=T_{\rm D}^{\rm FO}$, and the corresponding SM temperature is denoted as $T_{\rm FO}$. The dark freeze-out mechanism has been explored in the literature, for example, in Refs.~\cite{Carlson:1992fn,Cheung:2010gj,Chu:2011be,Bernal:2015ova}. One can define a quantity $\xi(T) \equiv T_{\rm D}/T$, that tracks the ratio of DS to visible sector temperature. For $s$-wave annihilation $\langle\sigma\cdot v \rangle=\sigma_0$, considering $2\to2$ process, one can then determine the asymptotic DM yield within the DS as~\cite{Kolb:1990vq},
\begin{equation}
   m_\psi Y_\psi^0 \simeq \frac{\xi(T_{\rm FO}) x_{\rm FO}^{\rm D}}{M_P\,\sigma_0 } \ ,
    \label{Eq:YpsiDFO}
\end{equation}
where $x_{\rm FO}^{\rm D} = m_\psi/T_{\rm FO}^{\rm D}$ and varies from ${\cal{O}}(20-30)$ as the DM mass varies from MeV to several TeV. For the given model, following Ref.~\cite{Cline:2013zca}, $\sigma_0 \simeq 1/\Lambda_{\rm dQCD}^2$. Therefore, to obtain the correct DM relic density
\begin{align}
   m_\psi Y_\psi^0  &\simeq 4.3 \times 10^{-10} {\,\rm GeV}  \left( \frac{x_{\rm FO}^{\rm D}}{35} \right) \left[ \frac{\xi(T_{\rm FO})}{0.3} \right] \left( \frac{\Lambda_{\rm dQCD}}{10 {\, \rm TeV}} \right)^2 \ .
   \label{Eq:relicAxion}
\end{align}
This points towards the dark QCD with dynamical scale $\Lambda_{\rm dQCD} \sim 10$ TeV, and therefore, the DM mass is $m_\psi \sim {\cal O}(50)$ TeV for $N_{D}=3$ as determined from Eq.~\eqref{Eq:DMMassAxion}. In this case, the DM behaves collisionless even at the dwarf galaxy scale as can be seen from Eq.~\eqref{Eq:sigmaDMDM}. 

Let us now turn to the question of stable DS relics other than the DM. For the current model, there are three dark pions $\pi_{\rm D}$, with mass $m_{\pi_{\rm D}} \simeq \sqrt{m_{\mathbf{q}} \Lambda_{\rm dQCD}}$, and can be determined from Eq.~\eqref{Eq:LambdadQCDrel} for a given $\Lambda_{\rm dQCD}$ and $\varepsilon$. Clearly, for $\Lambda \simeq 2$ meV, as required by the DE energy density, and $\Lambda_{\rm dQCD} \simeq 10$ TeV, as required by the dark freeze-out mechanism, $m_{\pi_{\rm D}} \lesssim 10^{-14}$ eV for $\varepsilon \gtrsim 10^{-10}$. Further, in the absence of dark electromagnetism and weak interaction, the pions are stable. They also do not decay to the DE axion $\varphi$ due to very suppressed mixing $\simeq \Lambda_{\rm dQCD}/(4 \pi f) \ll 1$. Hence, they behave as stable dark radiation and can be constrained by extra number of relativistic degrees of freedom $\Delta N_{\rm eff}$. If one assumes that the entropy is separately conserved in the visible and in the DS, then it is possible to write, $\xi\propto\left[\gss(T)/\gss^{\rm D}(T_{\rm D})\right]^{1/3}$, where $\gss$ and $\gss^{\rm D}$ are the number of relativistic degrees of freedom for entropy in the visible and the DS, respectively. As discussed in~\cite{Cheung:2010gj}, it is possible to constrain $\xi$ from the measurement of $ \DNeff=\frac{4}{7}\,\left(\frac{11}{4}\right)^{4/3}\,\gs^{\rm D}(T_{\rm CMB})\,\xi(T_{\rm CMB})^4$,
that takes care of the number of extra relativistic degrees of freedom beyond the SM at the time of CMB. Considering $\DNeff \lesssim 0.34$~\cite{Planck:2018vyg} from CMB observation, one finds, $\xi(T_{\rm CMB}) \lesssim 0.48$, for $\gs^D(T_{\rm CMB})=3$, as relevant for the three dark pions. Therefore, it is possible to tune $\Delta N_{\rm eff}$ by choosing an appropriate $\xi(T_{\rm CMB})$, consistent with the observations\footnote{An enhanced number of relativistic species (e.g., dark radiation) increases the early-Universe radiation density and expansion rate, advancing matter-radiation equality and reducing the sound horizon. This leads to a larger inferred Hubble constant, with one extra species raising \(H_0\) by about 7 km\,s\(^{-1}\)\,Mpc\(^{-1}\) in a flat Universe, enough to ease the Hubble tension~\cite{Riess:2016jrr}. The tension is effectively resolved for $\DNeff\sim 0.4$, though it reappears once BAO data is included~\cite{Bernal:2016gxb}. A detailed analysis of possible non-zero $\DNeff$ and its effect on the $H_0$, and $S_8$ tension, is beyond the scope of the present study.}. It is also possible to introduce portal operators for the dark pions to decay to SM particles, and then both the visible and DS can be in thermal equilibrium, and can introduce laboratory probe channels for the DM~\cite{Fujikura:2024jto}. We do not pursue this possibility here. In passing, we mention that in the present scenario, it is possible to have dark glueballs generated in the thermal bath. However, they will be unstable and decay into the dark pions, given that the dark pion mass is tiny. Further, if the confining phase transition of this dark QCD sector is first-order, which is expected for large $N_{\rm D}$, it could be associated with gravitational waves (GWs). It is expected that the scale $\Lambda_{\rm dQCD}$ would be responsible for setting the reheating temperature scale after this phase transition. Therefore, the gravitational waves would have a peak frequency in the range~\cite{Santos:2022hlx, Fujikura:2023lkn}
\begin{equation}
    f_{0} \simeq 10^{-3}  \left( \frac{\Lambda_{\rm dQCD}}{10 {\, \rm TeV}}\right) {\, \rm Hz} \ ,
    \label{Eq:fpeak}
\end{equation}
which is the relevant frequency range for LISA~\cite{Caprini:2019egz}, TianQin~\cite{TianQin:2020hid}, and 
Taiji~\cite{Hu:2017mde}.

\item [(ii)] {\it\underline{Asymmetric dark matter}:} As $\psi$ can be a Dirac fermion, it possesses a global conserved number symmetry $U(1)_{\rm D}$ similar to baryons in the visible sector. Therefore, it is quite natural if the DM relic density is set by a primordial particle-antiparticle number asymmetry similar to the visible sector. However, this requires an extension of the current setup by a portal operator that violates visible baryon number and $U(1)_{\rm D}$. The primordial asymmetry can be created in the visible sector, for \textit{e.g.,} via leptogenesis~\cite{Fukugita:1986hr} mechanism and can be reprocessed in the DM asymmetry, or the asymmetry can be created in the DS first, which is later shared with the visible sector. If the associated phase transition in the dark QCD sector is strongly supercooled, then any pre-existing asymmetry before the phase transition could be significantly diluted, and therefore, the latter possibility might be preferred in that case.

The asymmetric dark matter (ADM) framework suggests that the portal operator sharing the asymmetries was efficient in the early Universe but subsequently froze out~\cite{Kaplan:2009ag,Shelton:2010ta,Petraki:2013wwa,Zurek:2013wia}. Due to the portal interaction, usually the asymmetry shared between the DM and the visible sector are similar in magnitude, \textit{i.e.,} $n_{\psi}-n_{\bar{\psi}} \sim n_{\rm b}-n_{\bar{b}}$, where $b$ denotes SM baryon. Therefore,
\begin{equation}
    \frac{\Omega_{\rm DM}^{(0)}}{\Omega_{\rm B}^{(0)}} = \frac{m_\psi (n_\psi-n_{\bar{\psi}})}{m_p (n_b -n_{\bar b})} \simeq \frac{m_\psi}{m_p} \simeq 5.4 \ ,
    \label{Eq:omegaRatio}
\end{equation}
where $m_p$ is the mass of the proton. Hence, the DM abundance can be explained in the ADM framework if $m_\psi \simeq 5$ GeV. From Eq.~\eqref{Eq:DMMassAxion}, this suggests $\Lambda_{\rm dQCD} \simeq 3/N_{\rm D}$ GeV. As the SM electroweak sphalerons have frozen out at temperatures relevant for the DM formation, the following neutron portal can be used for sharing the asymmetry
\begin{equation}
    {\cal L}_n = \frac{1}{\Lambda_n^2} \widetilde{\psi} u_{R} d_{R} d_{R} \ ,
    \label{Eq:nportal}
\end{equation}
where $\widetilde{\psi}$ is any DS particle other than the DM and carries $U(1)_{\rm D}$ number, and $\Lambda_n$ is the effective cut-off scale for the neutron portal operator. Bound neutron decay is avoided kinematically if $\widetilde{\psi}$ is heavier than a neutron. Further, $\widetilde{\psi}$ needs to share the asymmetry inherited through the portal in Eq.~\eqref{Eq:nportal} to the DM $\psi$. In the current setup, $\widetilde{\psi}$ can be identified with the heavier baryon partner of $\psi$. When $\widetilde{\psi}$ decays to $\psi$, its asymmetry is shared partly with the DM. Here we have assumed that a primordial baryon asymmetry already existed in the visible sector via some mechanism. On the other hand, an asymmetry can be generated in the dark QCD sector by following Ref.~\cite{Fujikura:2024jto}, and can be shared with the visible sector. A portal can be introduced to make the dark pions decay before BBN. 
In this scenario, the dark and visible sectors come into thermal equilibrium due to the portal interactions, and the dark QCD confinement can trigger a phase transition in the DS that may explain the nano-Hz stochastic gravitational waves observed by pulsar timing arrays (see Ref.~\cite{Fujikura:2024jto} for more details). Moreover, $\psi$ is part of a strongly interacting sector with this given $\Lambda_{\rm dQCD}$, the DM self-interactions become important for evaluating properties of neutron stars~\cite {Kumar:2025yei}, and can explain the diverse rotation curves of spiral galaxies~\cite{Roberts:2024uyw}. 
\end{itemize}
We acknowledge that additional model-building efforts could yield alternative mechanisms for dark matter genesis within this framework. Nevertheless, our approach remains minimal and economical with respect to model parameters.
\section{Conclusions}
\label{eq:concl}
In this work, we have explored a consistent and minimal framework for an interacting DE-DM scenario that successfully explains the phantom crossing behavior of the evolving DE equation-of-state parameter, as indicated by recent DESI observations, without inducing any pathological features. We have introduced a dimension-5 effective operator involving the DE scalar and the SM photons, which accounts for the observed non-zero ICB angle. To explain the evolving DE behavior, we consider two well-motivated potentials for the DE scalar field: a polynomial potential, representative of tracker behavior in conventional quintessence models, and an axion-like potential naturally arising in strongly-coupled QCD-like dark sectors. We have provided a UV complete model for the latter. Furthermore, we have demonstrated how the DM can minimally obtain its relic density. Interestingly, the polynomial potential supports a superheavy freeze-in DM candidate with mass $\gtrsim 10^{10}\,\text{GeV}$ and requires a reheating temperature $\gtrsim 10^{12}\,\text{GeV}$, whereas the axion-like potential can yield a multi-TeV DM candidate for a dark freeze-out scenario, or a self-interacting GeV-scale DM candidate for an asymmetric DM model. The latter needs extension of the parameters, but can be accommodated naturally. They remain consistent with the combined constraints from DESI+CMB+DESY5 datasets within the $1\sigma$ level, along with the observed ICB angle.  
For the axion-type potential that emerges from a dark QCD sector, it is conceivable that there is an associated first-order phase transition in the DS. We have outlined that the dark freeze-out scenario may have associated gravitational waves with peak frequency in the range relevant for LISA~\cite{Caprini:2019egz}, TianQin~\cite{TianQin:2020hid}, and Taiji~\cite{Hu:2017mde}, while the asymmetric DM scenario suggests gravitational waves in the nano-Hz regime. A full analysis of the phase transition aspect of this model remains to be studied. In conclusion, our framework offers a unified and minimal approach that simultaneously addresses the DESI anomaly, the origin of cosmic birefringence, and the production of particle DM consistent with the observed relic abundance measured by Planck. We highlight that upcoming DESI observations, especially with improved sensitivity at high redshifts, can provide critical tests of, or potentially falsify, this scenario.
\acknowledgments
SG acknowledges fruitful discussions with Yuichiro Nakai, Yu-Cheng Qiu, Robert Shrock, and Hong-Yi Zhang.
\bibliography{Bibliography}
\bibliographystyle{JHEP}
\end{document}